\documentclass[]{aastex631}

\usepackage[utf8]{inputenc}
\usepackage[english]{babel}
\usepackage{array}
\usepackage{datatool}
\usepackage{soul} \usepackage{color} 
\usepackage[flushleft]{threeparttable}
\usepackage{amsmath}
\usepackage{float}
\usepackage{amssymb}
\usepackage{graphicx}

\usepackage{multirow}
\usepackage{longtable}

\begin{document}

\title{Hydroxyl Lines and Moonlight: a High Spectral Resolution Investigation of NIR skylines from Maunakea to guide NIR spectroscopic surveys}

\author[0000-0001-5123-6388]{Frederick Dauphin}
\affiliation{Space Telescope Science Institute, Baltimore, MD 21218}

\author[0000-0003-4030-3455]{Andreea Petric}
\affiliation{Space Telescope Science Institute, Baltimore, MD 21218}
\affiliation{William H. Miller III Department of Physics and Astronomy, Johns Hopkins University, Baltimore, MD 21218, USA}

\author[0000-0003-3506-5667]{Étienne Artigau}
\affiliation{Trottier Institute for Research on Exoplanets, Université de Montréal, Département de Physique, C.P. 6128 Succ. Centre-ville, Montréal,
    QC H3C 3J7, Canada}
\affiliation{Observatoire du Mont-M\'egantic, Universit\'e de Montr\'eal, D\'epartement de Physique, C.P. 6128 Succ. Centre-ville, Montr\'eal, QC H3C 3J7, Canada}

\author[0000-0002-4434-2307]{Andrew W. Stephens}
\affiliation{Gemini Observatory, NSF's NOIRLab, Hilo, HI, 96720}

\author[0000-0003-4166-4121]{Neil James Cook}
\affiliation{Trottier Institute for Research on Exoplanets, Université de Montréal, Département de Physique, C.P. 6128 Succ. Centre-ville, Montréal,
    QC H3C 3J7, Canada}

\author{Steven Businger}
\affiliation{University of Hawaii, Honolulu, HI 96822}

\author{Nicolas Flagey}
\affiliation{Space Telescope Science Institute, Baltimore, MD 21218}
\affiliation{Maunakea Spectroscopic Explorer Project Office, Canada-France-Hawaii-Telescope}

\author{Jennifer Marshall}
\affiliation{Texas A$\&$M University, College Station, TX 7784}
\affiliation{Maunakea Spectroscopic Explorer Project Office, Canada-France-Hawaii-Telescope}

\author{Michelle Ntampaka}
\affiliation{Space Telescope Science Institute, Baltimore, MD 21218}
\affiliation{William H. Miller III Department of Physics and Astronomy, Johns Hopkins University, Baltimore, MD 21218, USA}

\author[0000-0002-5269-6527]{Swara Ravindranath}
\affiliation{Astrophysics Science Division, NASA Goddard Space Flight Center, 8800 Greenbelt Road, Greenbelt, MD 20771, USA}
\affiliation{Center for Research and Exploration in Space Science and Technology II, Department of Physics, Catholic University of America, 620 Michigan Ave N.E., Washington DC 20064, USA}

\author{Laurie Rousseau-Nepton}
\affiliation{Canada-France-Hawaii Telescope, Waimea, HI 96743}
\affiliation{University of Toronto, Toronto, CA}
\affiliation{Dunlap Institute, Toronto, CA}

\begin{abstract}
Subtracting the changing sky contribution from the near-infrared (NIR) spectra of faint astronomical objects is challenging and crucial to a wide range of science cases such as estimating the velocity dispersions of dwarf galaxies, studying the gas dynamics in faint galaxies, measuring accurate redshifts, and any spectroscopic studies of faint targets. Since the sky background varies with time and location, NIR spectral observations, especially those employing fiber spectrometers and targeting extended sources, require frequent sky-only observations for calibration. However, sky subtraction can be optimized with sufficient a priori knowledge of the sky's variability. In this work, we explore how to optimize sky subtraction by analyzing 1075 high-resolution NIR spectra from the CFHT's SPIRou on Maunakea, and we estimate the variability of 481 hydroxyl (OH) lines. These spectra were collected during two sets of three nights dedicated to obtaining sky observations every five and a half minutes. During the first set, we observed how the Moon affects the NIR, which has not been accurately measured at these wavelengths. We suggest accounting for the Moon contribution at separation distances less than 10$^\circ$ when 1) reconstructing the sky using principal component analysis 2) observing targets at $YJHK$ mags fainter than $\sim$15 and 3) attempting a sky subtraction better than 1\%. We also identified 126 spectral doublets, or OH lines that split into at least two components, at SPIRou's resolution. In addition, we used Lomb-Scargle Periodograms and Gaussian process regression to estimate that most OH lines vary on similar timescales, which provides a valuable input for IR spectroscopic survey strategies. The \href{https://zenodo.org/records/13363061}{data} and \href{https://github.com/FDauphin/spirou-sky-subtraction}{code} developed for this study are publicly available.

\end{abstract}

\keywords{Night sky brightness(1112) ---  Time series analysis(1916) --- Calibration(2179) -- Infrared spectroscopy(2285)}

\section{Introduction} \label{intro}

Ground-based observations in the visible and the Near-InfraRed (NIR) are affected by the intensity, spatial, and spectral variations of the atmosphere. Sub-optimal sky subtraction impacts the uncertainty of the spectra extracted from the science targets and affects the final sensitivity achieved. Surveys targeting sources with fluxes similar to the background levels, and spectral features adjacent to sky lines are challenging \citep[e.g.,][]{soto16} and require adequate sky-subtraction strategies informed by appropriate knowledge of typical sky variations for the observing site. In this paper we present a phenomenological approach to characterize the IR skies, facilitated by six nights of high spectral resolution observations dedicated to measuring sky lines and continuum. 

Sky emissions in the visible are dominated by atmospheric scattering, while sky emissions in the NIR are dominated by airglow emission, water, $O_2$, and $N_2$ lines. The NIR airglow originates in OH radicals made by reactions between hydrogen and ozone from a layer about 9 km thick at an altitude of about 90 km \citep{elib08}.  

Several studies suggest that 0.5-1\% accuracy can be reached with low resolution (R$<~5000$) spectrographs for faint (23\,mag in the $J$ and $H$, NIR bandpasses) targets with massively multiplexed systems and wide fields of view \citep{sp2010, Zhang2016, zap2016, rod2010, pue2010}. This can be achieved by employing principal component analysis (PCA) methods or using PCA together with 2D modeling of the sky obtained from fibers allocated to the sky and improved by a well-constrained, a priori knowledge of the sky variations.

For Southern skies, European Southern Observatory (ESO) facilities, benefited from efforts combining decades-old atmospheric conditions with insight from physical models of the atmosphere \citep{2012_noll, 2014_noll, noll23}. 

{\it ESO SkyCalc} can be used by observers aiming to optimize photometric and spectroscopic observations in the optical and NIR, especially those of faint targets, or faint lines adjacent to sky lines. The {\it ESO SkyCalc} tool illustrates the importance of using physical models for sky subtraction. 

In this paper, we present a potential first step toward the development of an {\it ESO SkyCalc} like model for Maunakea. We present an unique set of high-resolution, high-cadence, NIR sky spectra that can be used to develop guidelines for ongoing and future spectroscopic surveys (see also \citet{buton2013}). We also illustrate how time series analysis of the temporal variations on NIR sky spectra can be used to increase the efficiency of NIR sky observations.

The primary goal of this work is to present guidelines for ongoing and upcoming IR spectroscopic surveys regarding the time between science observations and sky frames as well as distance to the Moon. These results have a significant bearing on survey efficiency and observing strategy. Specifically, survey design includes developing scheduling strategies that optimize sky exposures as a function of the targets' emission line locations, the targets' magnitudes, and the atmospheric conditions.

The paper is organized as follows. In Section \ref{sec:dataset}, we present SPIRou and our observations. In Section \ref{sec:methods}, we describe our methods for fitting the hydroxyl lines to our flux model and measuring their variability using Gaussian process regression and the Lomb-Scargle Periodogram. In Section \ref{sec:results}, we describe our results. We summarize spectroscopic survey guidelines in Section \ref{sec:discuss}, and conclude our work in Section \ref{sec:conclude}. Finally, in the Appendix, we discuss data availability and related analyses beyond the scope of the text.

\section{Spectroscopic Sky Data Set using SPIRou} \label{sec:dataset}

\subsection{SPIRou: The SpectroPolarimètre Infra-Rouge} \label{sec:SPIRou}

\normalsize{
The spectra used for our investigation were obtained with SPIRou (SpectroPolarimètre Infra-Rouge; \citet{2018_donati}), a high spectral resolution (R $\sim$ 75,000) NIR spectrograph at the Canada-France-Hawaii Telescope's (CFHT) 3.6-m telescope on Maunakea.  SPIRou is a high-resolution near-infrared spectropolarimeter and high-precision velocimeter that saw first light in April 2018. Its primary science objectives are to characterize and detect habitable exoplanets orbiting low-mass stars, and understand how magnetic fields affect star and planet formation \citep{2015_moutou,2014_artigau, 2018_donati}. 

In total, we collected 1075 sky observations, which spanned from July 28$^{\rm th}$, 2018 to January 10$^{\rm th}$, 2022, or approximately 3.5\,years. The observations were also unevenly spaced, ranging from one minute to twelve weeks between observations. In December 2019 and January 2020, the observatory experienced technical problems that prevented pointing, guiding, or any motion of the telescope thus stopping all science operations. We decided to use the time to observe the night sky with SPIRou. Due to the technical issue, all sky observations during those nights where taken at zenith. These observations included two sets of three days dedicated to sky measurement where each day, a sky spectrum was observed approximately every 5.5\,minutes for 12\,hours. These days occurred on December 14$^{\rm th}$, 15$^{\rm th}$, and 16$^{\rm th}$ of 2019 and January 22$^{\rm nd}$, 23$^{\rm rd}$, and 25$^{\rm th}$ 2020.  For the rest of this work, we refer to the first three days as Event 1 and the last three days as Event 2. In Event 1, the five-minute observing cadence ceased for about 1.5 hours, causing a temporal gap on the first day. In addition, unplanned at the time since the telescope was not tracking, the Moon passed in close of the fibers during Event 1. These anomalous data were removed in the analysis of OH emission, but these data provide the first strong constraint of thermal continuum contribution of the Moon at NIR wavelengths on Maunakea, complementing the \cite{2019A&A...624A..39J} study of scattered moonlight using the Very Large Telescope (VLT). Event 1 contained 348 spectra (32\% of the 1075 observations) and Event 2 contained 372 spectra (35\% of the 1075 observations). Figure \ref{fig:obs_coords} shows where the observations were taken on the sky (Altitude, Azimuth in degrees): Event 1 was split between $(89.925^\circ, 170.269^\circ)$ and $(89.925^\circ, 179.973^\circ)$, while Event 2 was only observed at $(89.925^\circ, 170.286^\circ)$.

Similar to other NIR detectors, SPIRou is susceptible to persistence \citep{2018_artigau}. High-flux exposures leave behind spectral features for up to several hours, marking the detector with a spurious background signal. Persistence naturally decays over the night. We describe our mitigation against persistence later in Section \ref{sec:optimize}.

The 2D raw sky spectra (see Figure \ref{fig:raw_data}) were reduced to 1D calibrated spectra (see Figure \ref{fig:example_spec}) using APERO: A PipelinE to Reduce Observations (version 0.7.228) \citep{2022_cook}. These 1D extracted and flat fielded sky spectra were supplied with a wavelength solution given in constant velocity bins. The pixel wavelengths were calibrated using a UNe hollow-cathode lamp and a Fabry-Pérot étalon to an internal error of $\sim0.15$ m/s \citep{2021_hobson}. The wavelength range of the spectra was $0.965-2.500{\mu}m$ containing 285\,377 wavelength bins resampled on a uniform wavelength grid with a step of 1 km/s/pixel, and since the pixels were constant in velocity, the change in wavelength increased from $3\times10^{-6}-8\times10^{-6}{\mu}m$ per pixel. Negative flux measurements were most likely due to poor dark-current extraction, bad pixel inclusion, and random negative excursions of the readout noise \citep{apero}. All observations were affected by a steep black body curve starting at $2.1{\mu}m$, which was caused by thermal emission. The exposure times were 300.884 seconds for both Events 1 and 2. Although the airglow intensity depends on the zenith angle, we ignored correcting for the zenith since there was minimal angle variation for Events 1 and 2.

\begin{figure}[t]
\centering
\includegraphics[scale=0.24]{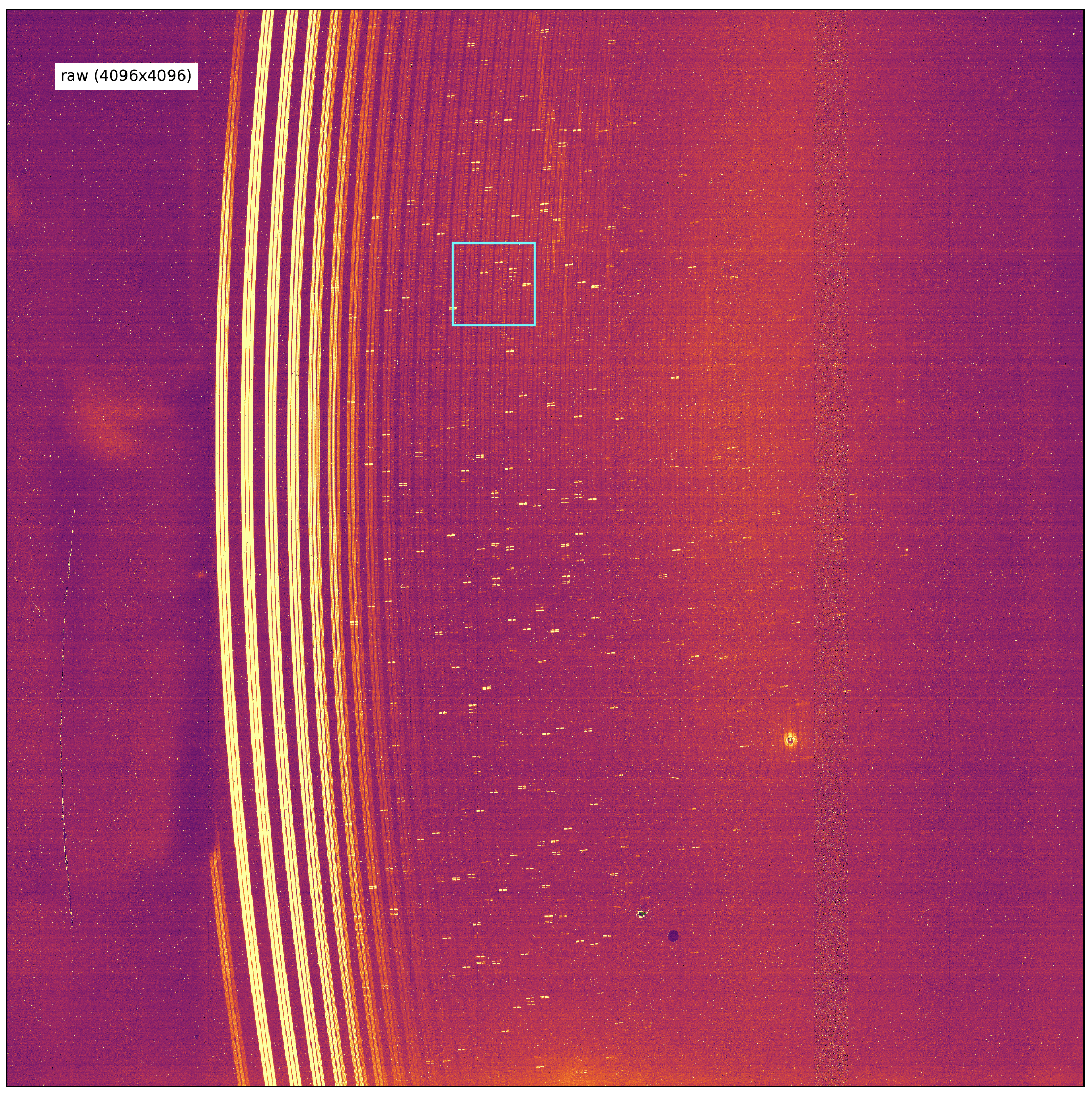}
\centering
\includegraphics[scale=0.24]{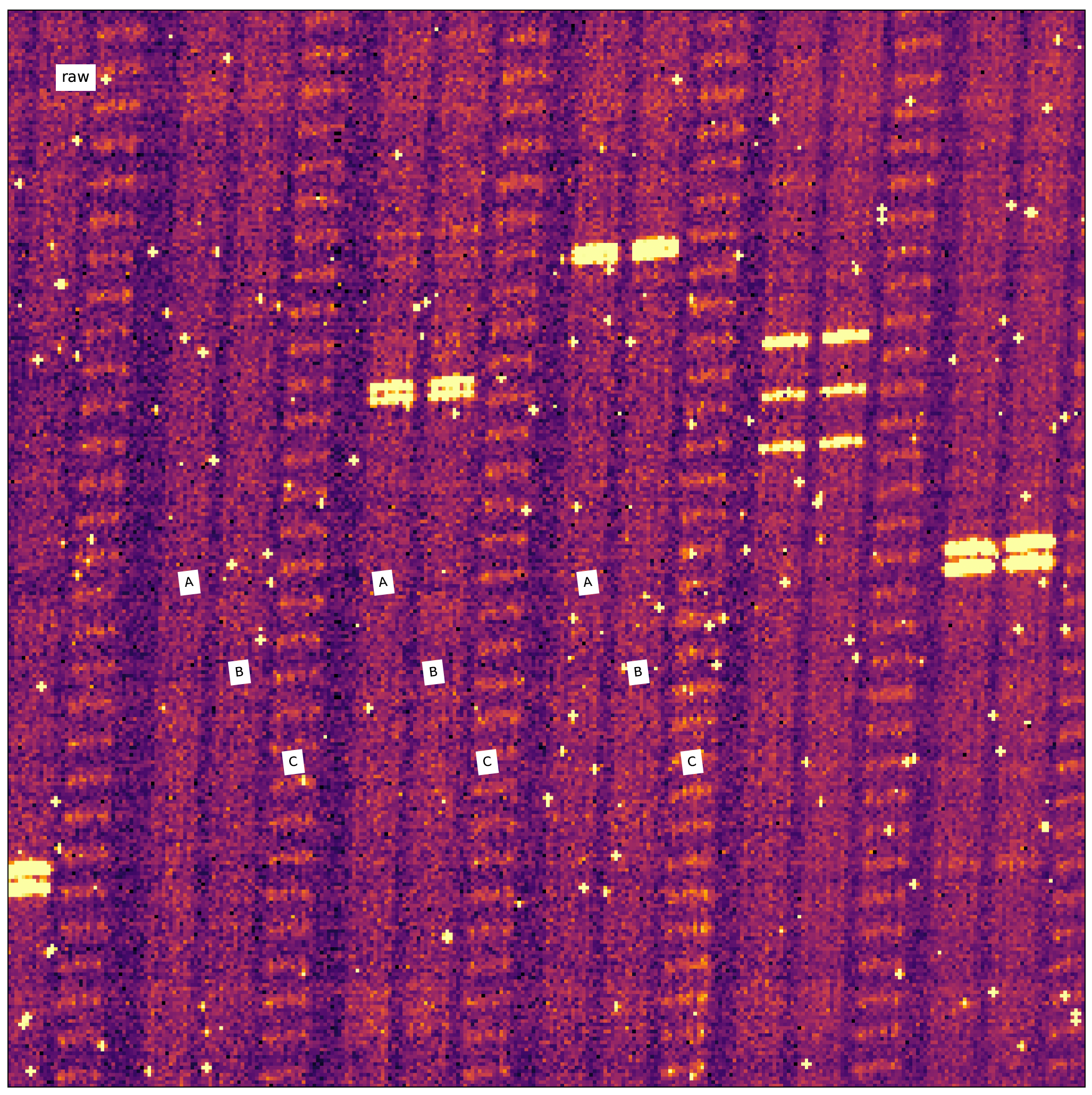}
\caption{A raw two-dimensional SPIRou sky spectrum. Left: A full-frame image with wavelength decreasing from left ($K$ band) to right ($Y$ band). The slicer shape, tilt, and curvature of the 49 echelle orders were corrected in calibration. Right: A zoom-in image of the region boxed in cyan on the left. The A, B, and C fibers correspond to the two science and one reference fiber, respectively.}\label{fig:raw_data}
\end{figure}

\begin{figure}[t]
\centering
\includegraphics[width=\linewidth]{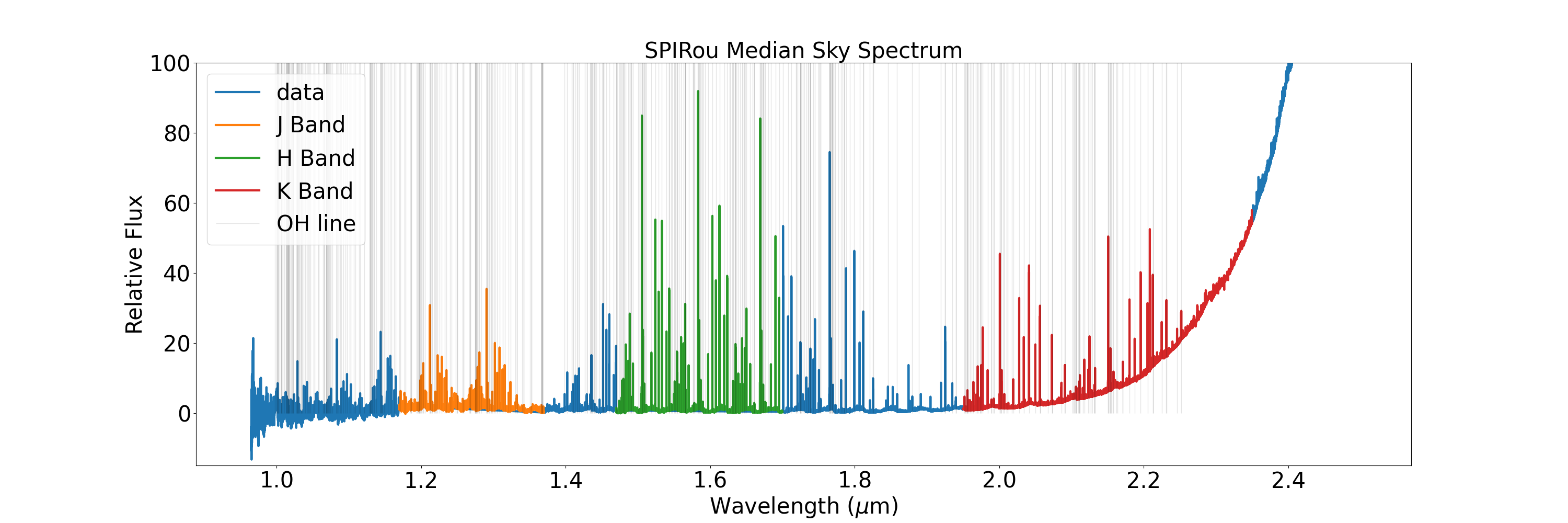}
\caption{Median stack of our calibrated SPIRou sky spectra. The flux is normalized by the spectrum's median. The $J$, $H$, and $K$ bands are shown in orange, green, and red, respectively. The 481 lines we fit as part of our analysis are shown in light gray. The black body curve, originating from the thermal emission of the telescope, the front-end module, and the optical fiber train, is also visible starting around $2.1{\mu}m$. The figure illustrates the multitude of sky lines in the NIR regime and also the need for high-resolution site-specific spectra to accurately estimate their variability and correlation.}
\label{fig:example_spec}
\end{figure}

\subsection{Hydroxyl (OH) Lines} \label{sec:oh_lines}

Hydroxyl emission lines from OH molecules in the Earth's atmosphere are abundant and prominent in the infrared, which heavily contaminates ground-based spectral observations \citep[e.g., ][]{bland, oli2015, oli2013}. OH are significant contributors to the sky emission in the NIR \footnote{$O_2$ are also significant contributors, but here we only focus on OH}. We used a set of 481 hydroxyl lines published by \cite{2000_rousselot}. That work presented a medium resolution spectra of OH sky emission between $0.997-2.25{\mu}m$ and $R\sim8000$. \citet{2000_rousselot} presented vacuum wavelengths, which we used, along with air wavelengths and line identifications. All of the OH lines were well aligned with our spectra, indicating similar wavelength calibration levels between our studies and theirs.

}

\section{Methods: Measuring OH Flux Contribution and Variability} \label{sec:methods}

In this section, we describe the four main components of our methods: our flux model, fitting and optimization methods (Section \ref{sec:optimize}), total flux contribution measurements (Section \ref{sec:integration_metrics}), and variability measurements (Section \ref{sec:lsp}). Each spectrum was normalized by the exposure time.
We tracked a few overall spectra metadata, such as exposure time, mid-observation MJD, airmass, spectrum median, $J$ band ($1.17-1.37{\mu}m$) median, $H$ band ($1.47-1.70{\mu}m$) median, $K$ band ($1.95-2.35{\mu}m$) median, and the ratio of the total $K$ band flux to the total $J$ band flux.

\subsection{Flux Model, Fitting, and Optimization} \label{sec:optimize}

\begin{figure}[t]
\centering
\includegraphics[scale=0.26]{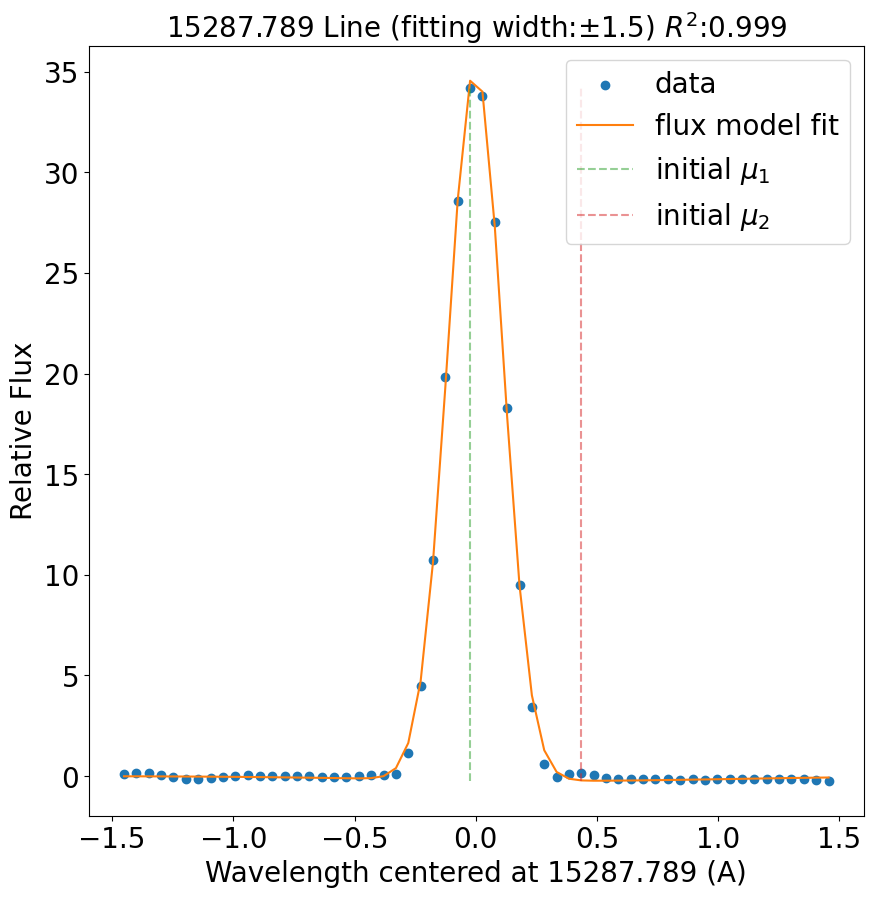}
\centering
\includegraphics[scale=0.26]{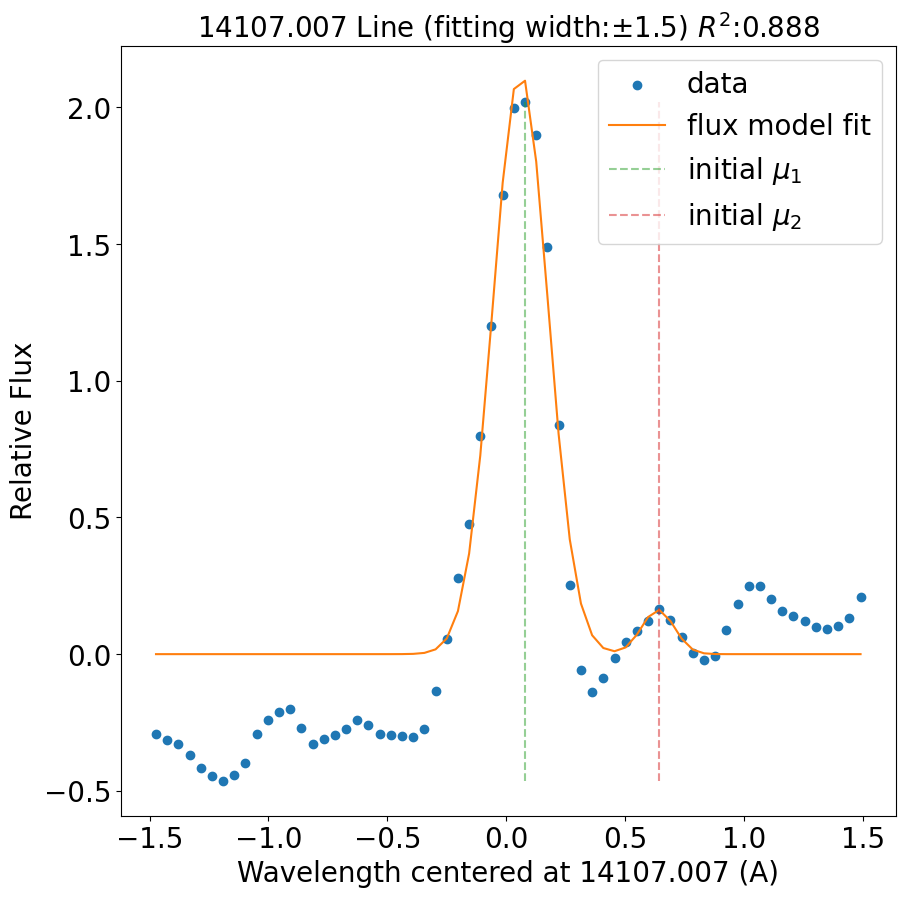}
\centering
\includegraphics[scale=0.26]{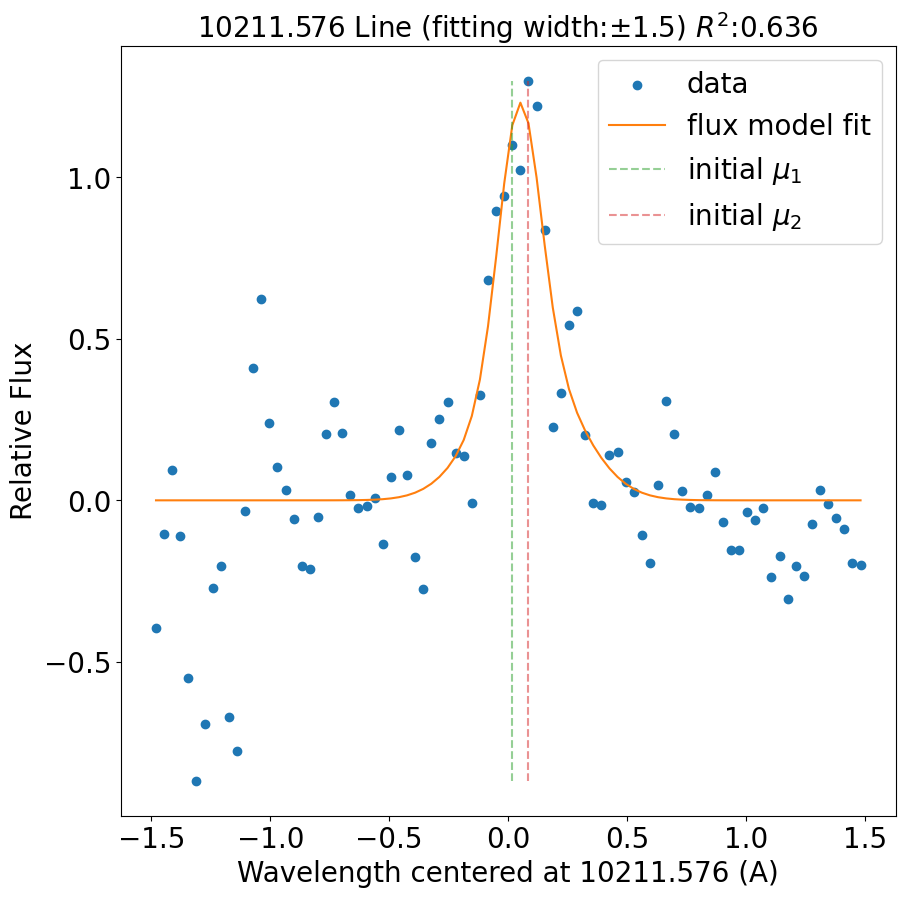}
\centering
\includegraphics[scale=0.26]{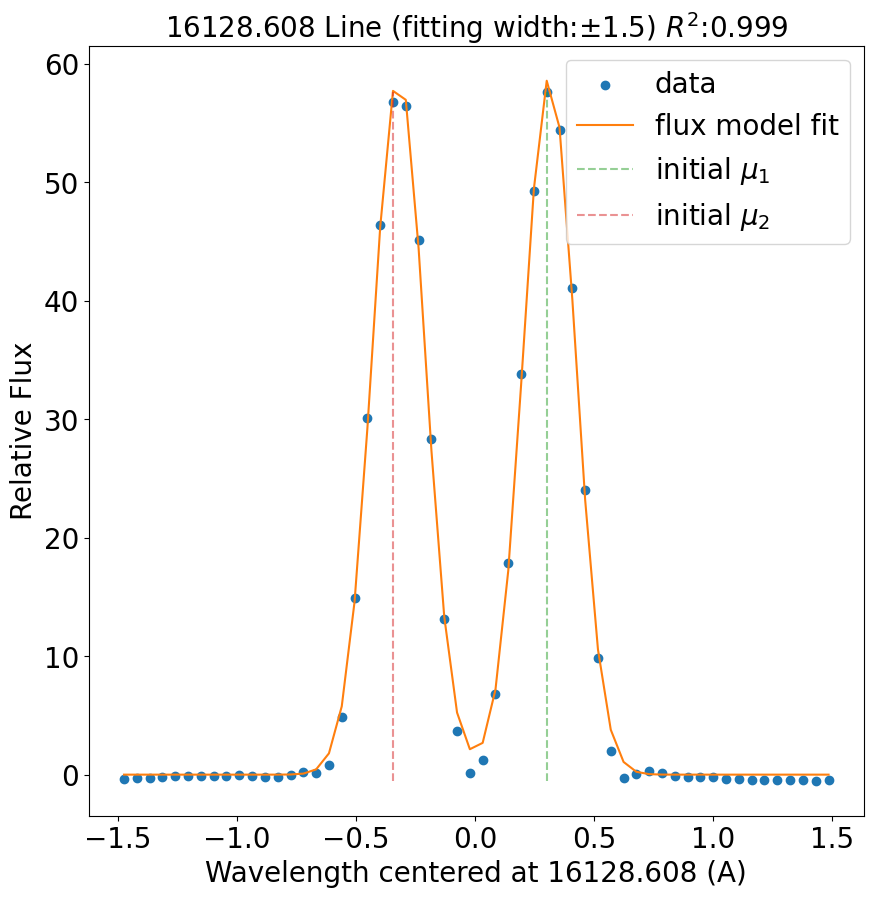}
\centering
\includegraphics[scale=0.26]{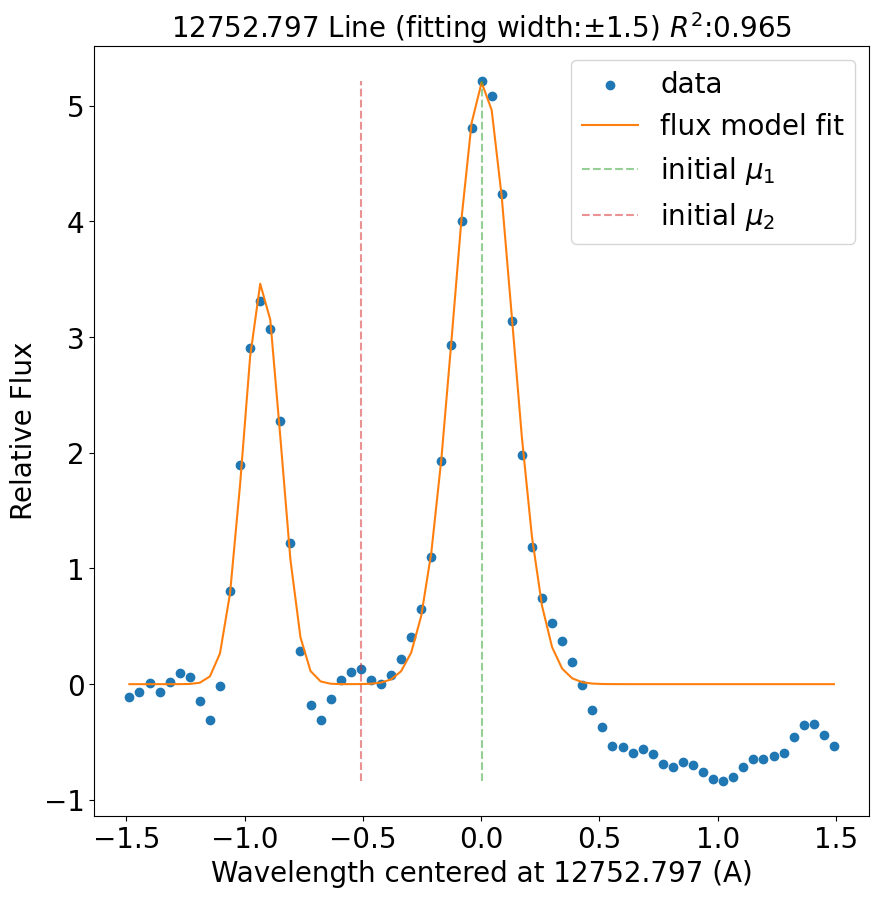}
\centering
\includegraphics[scale=0.26]{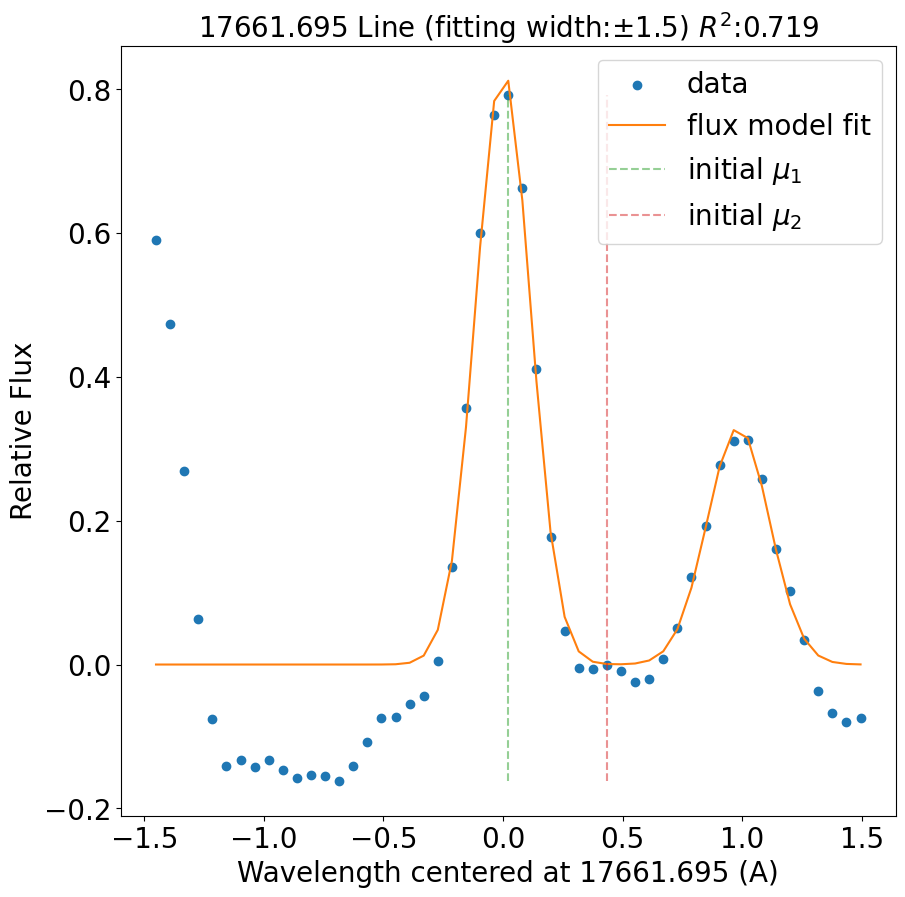}
\caption{Examples of flux model fits for singlet (top) and doublet (bottom) profiles. The left shows a flat continuum, which is well-fit by a Gaussian line emission on top of a constant continuum model.  The middle column shows a non-flat continuum. The right column shows poor fits, indicated by large discrepancies between the flux model fit and the data. The initial wavelengths for $\mu_1$ and $\mu_2$ are shown as dashed green and red lines, respectively. The flux was normalized by the median of the spectrum used for fitting. $R^2$ was the coefficient of determination (i.e. square of the Pearson correlation). Our model was versatile enough to fit singlet and doublet profiles on a smooth continuum, but struggled against noisy (top-right) and multi-featured local spectra (bottom-right).}
\label{fig:model_fits}
\end{figure}

\normalsize{

The primary component of our OH line flux model was a Gaussian as a function of wavelength ($\lambda$) with parameters amplitude ($A_i$), line center ($\mu_i$), and line width ($\sigma_i$). Our model incorporated two Gaussian functions instead of one to account for doublets and other non-flat background components \citep{bland,oli2015,sh85}, as follows in Equation \ref{eq:3}.

\begin{equation}
\label{eq:3}
F_{flux,\lambda}(\lambda) = F_{G,\lambda}(\lambda; A_1, \mu_1, \sigma_1) + F_{G,\lambda}(\lambda; A_2, \mu_2, \sigma_2)
\end{equation}

where the equation estimated the flux contribution from the OH line and line splitting at $\lambda$, respectively.

Before fitting, each spectrum was high-passed using a splined running median as a low-pass filter. The low-pass filter removed any empirical waveband continuum due to known systematic errors, e.g., in the farther infrared or persistence, and any unknown background noise. Since nonlinear least-squares optimizers require strong initial values to find best-fit parameters, we chose our initial values before fitting by performing the following:

\begin{enumerate}

    \item \textbf{Retrieve a median-subtracted local spectrum centered on the OH line $\pm1.5\times10^{-4}{\mu}m$.} These bounds best isolated our lines from other sources, and included a sufficient amount of data points to fit (40-90 for redder-bluer wavelengths). NaNs were also excluded from the local spectrum.

    \item \textbf{Smooth the local spectrum with a Gaussian filter ($N(0, 2)$) to remove low SNR local maxima.} A normal distribution with $2\sigma$ empirically worked best for our data processing needs. 

    \item \textbf{Choose the closest local maxima to the OH line to be the initial $\mu_1$, and the corresponding flux to be the initial $A_1$.}

    \item \textbf{Set $\mu_2$ and $A_2$ based on the next highest local maxima if one exists within $\pm1.5\mathring{A}$ of the OH line.}

\end{enumerate}

$\sigma_1$ and $\sigma_2$ were initially set to $10^{-5}{\mu}m$ since most lines' widths were within that order of magnitude. Finally, we fit the OH lines using traditional least-squares optimization up to 10\,000 optimization steps. If no best-fit parameters were found or if both Gaussians were fit outside of the local spectrum, all parameters were changed to NaN.

Figure \ref{fig:model_fits} demonstrates fits where the model performed to varying degrees in order to understand the conditions for success and failure. The model was able to fit a diverse set of emission lines due to its several parameters.

}

\subsection{Methods: Flux Integration and Recorded Metrics} \label{sec:integration_metrics}

\normalsize{

To estimate the true OH contribution of the fitted flux $\hat{F}_{flux, \lambda}$ and the observed flux $F_{flux}$, we subtracted the second Gaussian $F_{G,\lambda,background}$ if the peaks differed by an order of magnitude as the dimmer component was most likely fit to the background and was not fitting a doublet:

\begin{equation}
\label{eq:4b}
\hat{F}_{flux,\lambda}(\lambda) = F_{fit,\lambda}(\lambda)
- F_{G,\lambda,background}(\lambda)
\end{equation}

\begin{equation}
\label{eq:5b}
F_{flux,\lambda}(\lambda) = F_{observed,\lambda}(\lambda) - F_{G,\lambda,background}(\lambda)
\end{equation}

where $F_{fit}$ was the flux fit by our model and $F_{observed}$ was the observed flux. Next, we integrated the fluxes across the entire local spectrum to calculate the total flux contribution of each line, as follows in Equation \ref{eq:6} and \ref{eq:7}:

\begin{equation}
\label{eq:6}
\hat{F}_{flux}(\lambda) = \int_{}^{} \! \hat{F}_{flux,\lambda}(\lambda) \, \mathrm{d}\lambda
\end{equation}

\begin{equation}
\label{eq:7}
F_{flux}(\lambda) = \int_{}^{} \! F_{flux,\lambda}(\lambda) \, \mathrm{d}\lambda
\end{equation}

We recorded several metrics to evaluate our fits: local spectrum median, local spectrum median absolute deviation ($MAD$), relative error ($\eta$), and variance, which we refer to as mean square error ($MSE_{F}$) to differentiate from the line widths), Pearson correlation ($R_{pearson}$), Spearman correlation ($R_{spearman}$, which is $R_{pearson}$ for the rank of variables), and its p-value as follows in Equations \ref{eq:8}-\ref{eq:11}:

\begin{equation}
\label{eq:8}
MAD = median(|F_{flux,\lambda}(\lambda) - \tilde{F}_{flux,\lambda}(\lambda)|)
\end{equation}

\begin{equation}
\label{eq:9}
\eta = \frac{|F_{flux}(\lambda) - \hat{F}_{flux}(\lambda)|}{F_{flux}(\lambda)}
\end{equation}

\begin{equation}
\label{eq:10}
MSE_F = \frac{1}{n} \sum_{i=1}^{n} (F_{flux,\lambda}(\lambda_i) - \hat{F}_{flux,\lambda}(\lambda_i)) ^ 2
\end{equation}

\begin{equation}
\label{eq:11}
R_{Pearson} = \frac{cov(F_{flux,\lambda}(\lambda_i), \hat{F}_{flux,\lambda}(\lambda_i))}{\sigma_{F_{flux,\lambda}(\lambda_i)} \sigma_{\hat{F}_{flux,\lambda}(\lambda_i)}}
\end{equation}

where $\lambda_i$ is a wavelength in the emission profile and $n$ is the number of samples in the emission profile (e.g. 40-90 for redder-bluer wavelengths)

}

\subsection{Methods: Short-term Flux Variability: Gaussian Proccess Regression} \label{sec:gpr}

We constructed a time series of the total flux contributions of each OH line. With these time series, we can a.) estimate the variability of OH, and b.) interpolate between gaps to predict flux with unknown measurements. Gaussian process regression (GPR) is an extremely versatile technique that can approximate a diverse set of smoothly varying functions \citep{2006_gp, 2023_aigrain}. They are built using covariance functions (or kernels). These covariance functions are used to calculate a covariance matrix based on the prior data points. One can draw samples from a multinormal distribution with a mean function (usually the data set's mean) and the covariance matrix to sample functions. GPRs are conditioned using training data, and optimized by maximizing log-likelihood. Once converged, these models can interpolate and extrapolate functions \citep{2006_gp, 2023_aigrain}.

Our covariance function was the squared exponential kernel (also known as the radial basis function) with a white noise kernel, as follows in Equation \ref{eq:rbf}:

\begin{equation}
\label{eq:rbf}
k(t_i, t_j) = A^2 exp(-\frac{(t_i - t_j)^2}{2l^2}) + \sigma^2
\end{equation}

where $A$ was the amplitude, $t_i$ and $t_j$ were the observational times of our spectra, $l$ was the length scale, which determines how wide of a distribution of points affects each other, and $\sigma$ was the estimated variance of the data. The GPR model was fitted on the residual to line intensity subtracted from a linear fit of amplitude against time through the entire sequence. Here, $A$ is therefore a measure of the variability amplitude, not the mean flux ($\mu$). The fractional change of a line with time is therefore $A/\mu$ (see Section~\ref{section:gpr}).  The white noise kernel helped fit additional noise in our data. In addition, we optimized parameters with the Markov Chain Monte Carlo (MCMC) method implemented in the Python package \texttt{emcee} to estimate the uncertainties in our parameters \citep{2013_foreman, 2019_speagle}. Our prior for $A$ and $\sigma$ was a log uniform distribution bounded by 0.01 and 100 times the standard deviation of the OH flux. Our prior for $l$ was a log uniform distribution bounded by 10 minutes and 1 day. It is important to note that a small amplitude-to-white-noise ratio ($A/\sigma$) indicates that most of the variance in the data is attributed to uncorrelated scatter and therefore that we have a poor ability to sample short-term time-correlations, which in turn biases toward higher $l$ values. One therefore expects the best constraints on $l$ to be obtained for high $A/\sigma$ time series.

\subsection{Long-term Flux Variability: Lomb-Scargle Periodogram} \label{sec:lsp}

\normalsize{

Traditionally, a Fourier transform would be adequate, but Fourier transforms assume evenly sampled data. Since our data was uneven, we used the Lomb-Scargle Periodogram (LSP), which produces a power spectrum for unevenly sampled data \citep{1976_lomb,1982_scargle}. Although we did not expect a truly periodic behavior as the OH line intensities were a combination of density waves and de-excitation timescales, LSP was still a solid baseline for time series analysis, specifically in decomposition and forecasting. The LSP equations were stated as follows \citep{2018_vanderplas}:

\begin{figure}[t]
\centering
\includegraphics[scale=0.4]{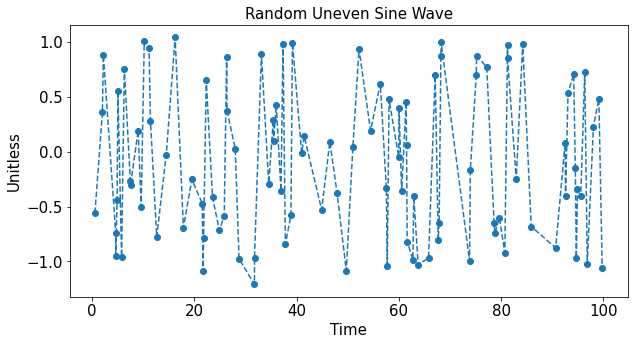}
\centering
\includegraphics[scale=0.4]{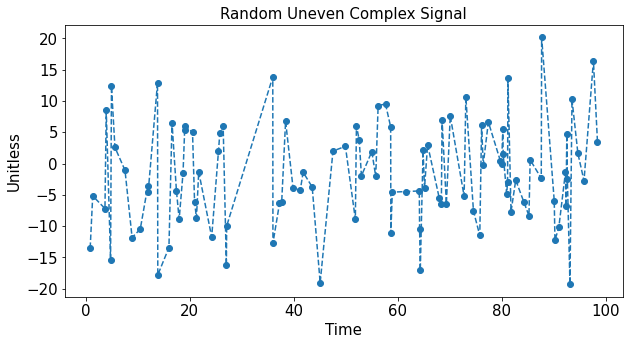}
\centering
\includegraphics[scale=0.375]{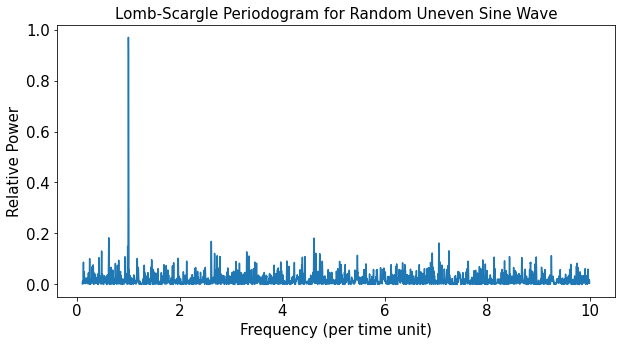}
\centering
\includegraphics[scale=0.375]{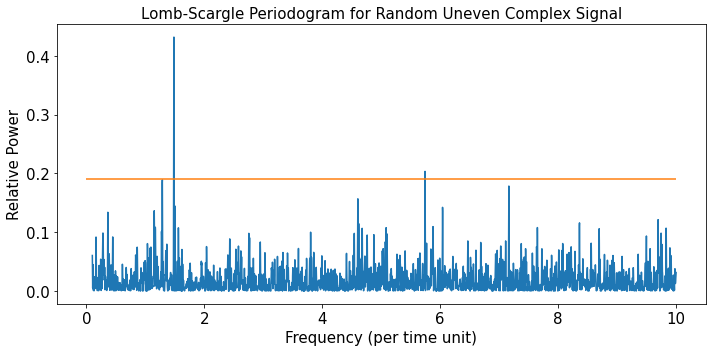}
\centering
\includegraphics[scale=0.375]{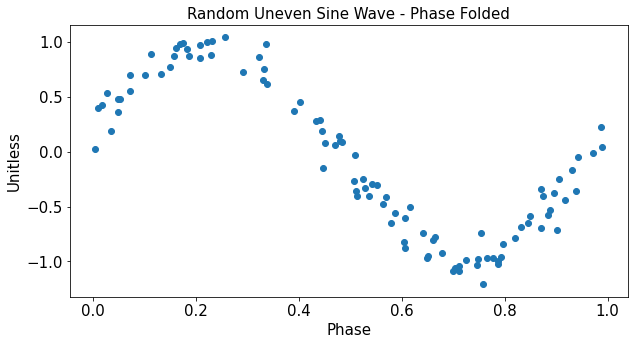}
\centering
\includegraphics[scale=0.375]{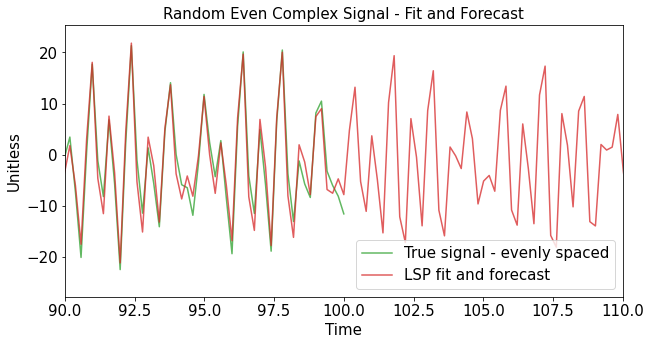}
\caption{This figure aims to present an overview of the Lomb-Scargle Periodogram (LSP) method and motivate why we use the method for our analysis. We show LSP examples on a noisy ($N(\mu=0,\sigma=0.1)$) sine ($f=1$) wave (left) and a noisy ($N(\mu=0,\sigma=5)$) complex wave (right), which was the sum of three different sine waves with random amplitudes, oscillation frequencies ($f=1.288, 1.485, 5.741$), phase shifts, and offsets. The top row shows the signals with random unevenly spaced time sampling, where visually noticeable periodic patterns were difficult to detect. The second row shows the power spectrum of each signal. The sine wave peaked at $f=0.9996$, and the complex wave peaked at four different locations: three true $f=1.287, 1.484, 5.742$ (at or above the orange line), and one false $f=7.165$ (below the orange line). The sine wave was phase folded on the bottom left to show one clean period. On the bottom right, the LSP fit at evenly spaced time samples (red) matched closely with the true signal at evenly spaced time samples (green) with a $R_{Pearson}^2=0.919$. The forecast after $t=100$ also followed similar periodic patterns as the true signal. The LSP fit was originally offset by -1.463 from the true signal so by subtracting the offset from the LSP fit, we retain a centered reconstruction. While the concept of LSP is well known in astronomy \citep{2018_vanderplas} we provide this example to motivate our analysis.}
\label{fig:lsp_example}
\end{figure}

\begin{equation}
\label{eq:12}
\tau = \frac{1}{4{\pi}f} \mathrm{tan}^{-1} \bigg( \sum_{n} \mathrm{sin}(4{\pi}ft_n) \Big/ 
\sum_{n} \mathrm{cos}(4{\pi}ft_n) \bigg)
\end{equation}

where $\tau$ ensured a time-shift invariance, $f$ was the oscillation frequency of the signal, and $t_n$ was the observing time of the signal.

\begin{equation}
\label{eq:13}
P_{LS,\mathrm{cos}}(f) = \bigg( \sum_{n} g_n \mathrm{cos}(2{\pi}f[t_n - \tau]) \bigg) ^ 2 \bigg/
\sum_{n} \mathrm{cos}^2(2{\pi}f[t_n - \tau])
\end{equation}

\begin{equation}
\label{eq:14}
P_{LS,\mathrm{sin}}(f) = \bigg( \sum_{n} g_n \mathrm{sin}(2{\pi}f[t_n - \tau]) \bigg) ^ 2 \bigg/
\sum_{n} \mathrm{sin}^2(2{\pi}f[t_n - \tau])
\end{equation}

where $P_{LS,\mathrm{cos}}$ was the cosine contribution to the power spectrum, $P_{LS,\mathrm{sin}}$ was the sine contribution to the power spectrum, and $g_n$ was a finite discrete analog to a true indefinitely long and continuous signal, acting as a weight term for every single frequency sinusoidal. Thus follows:

\begin{equation}
\label{eq:15}
P_{LS}(f) = \frac{1}{2} (P_{LS,\mathrm{cos}}(f) + P_{LS,\mathrm{sin}}(f))
\end{equation}

where $P_{LS}$ is the average of the cosine and sine contributions. Two examples are demonstrated in Figure \ref{fig:lsp_example}.

We implemented LSP as follows:

\begin{enumerate}

    \item Center the time axis so 0 is the minimum MJD. 
    
    \item Mask out any NaNs from the fits and remove fits with $\eta<15\%$ to ensure only good measurements were kept.
    
    \item Produce a power spectrum using LSP between periods of 1 year per cycle to 4 minutes per cycle ($1/365$ cycles per day to 360 cycles per day). The frequency bins were automatically chosen by the \texttt{astropy} algorithm \citep{2018_astropy}.
    
    \item Detect the top five most dominant frequencies, which sufficiently capture most of the variation in the time series.
    
\end{enumerate}

We reconstructed the time series using the five most dominant frequencies. Since the \texttt{astropy} implementation only handled singular frequency reconstruction, we reconstructed the time series by adding five singular frequency models (similar to the bottom right of Figure \ref{fig:lsp_example}). To center the reconstruction, we estimated the mean difference between the reconstruction and total flux contributions and subtracted the difference from the reconstruction. Finally, we calculated $\eta$, $MSE$, $R_{Pearson}$, and $R_{Spearman}$ between the centered reconstruction and the total flux contribution.

}

\clearpage
\begin{figure}[!htbp]
\centering
\includegraphics[width=\linewidth]{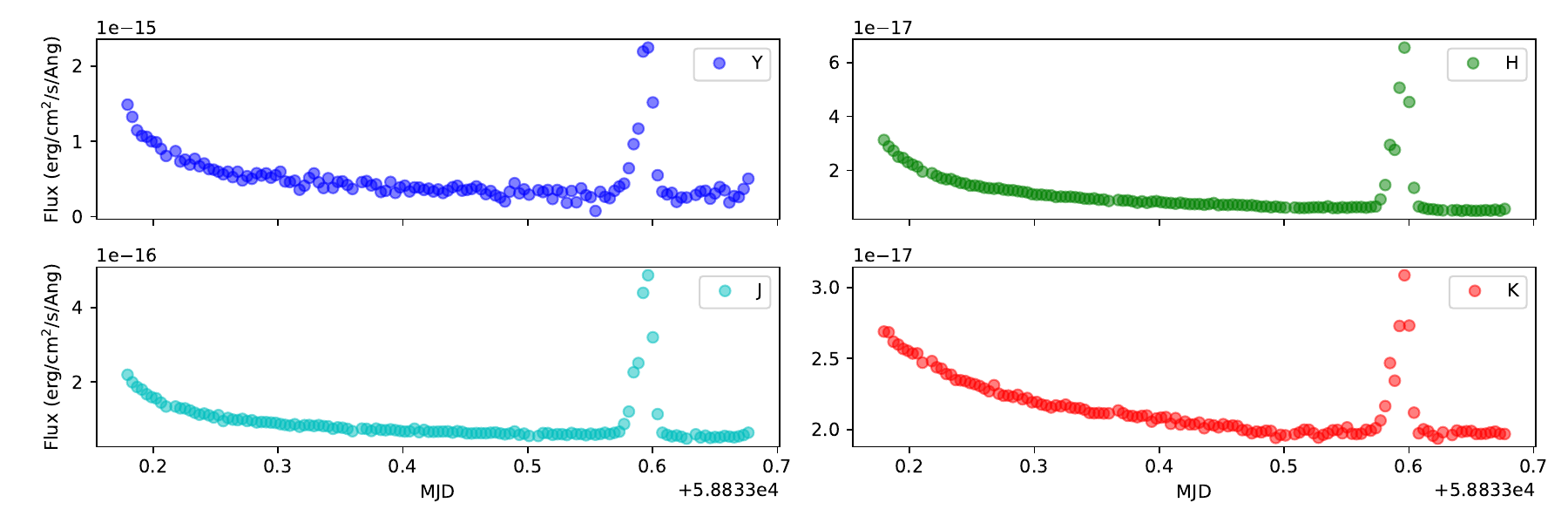}
\caption{Median sky backgrounds in $YJHK$ bands. One clearly sees a low-level persistence decay from the evening calibrations within each band. The peak at the end of the night is due to the Moon continuum contribution as it serendipitously passes within $<2^\circ$ of SPIRou's fiber.}
\label{fig:raw_bgnd}
\end{figure}

\begin{figure}[!htbp]
\centering
\includegraphics[width=\linewidth]{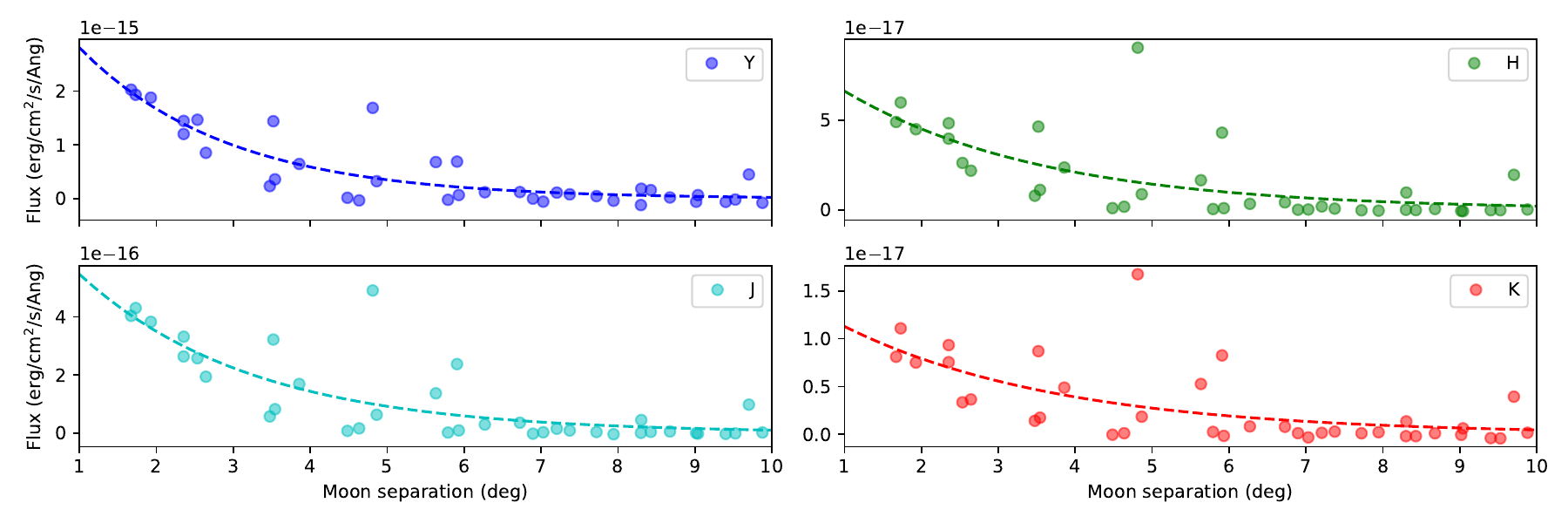}
\caption{Continuum background level contribution as a function of the angular separation of the fiber to the Moon. A simple exponential decay was fitted to each band dataset. All 3 nights with the Moon passage were included in the fit and we did not correct for the Moon phase change between dates.}
\label{fig:raw_bgnd2}
\end{figure}

\begin{figure}[!h]
\centering
\includegraphics[width=\linewidth]{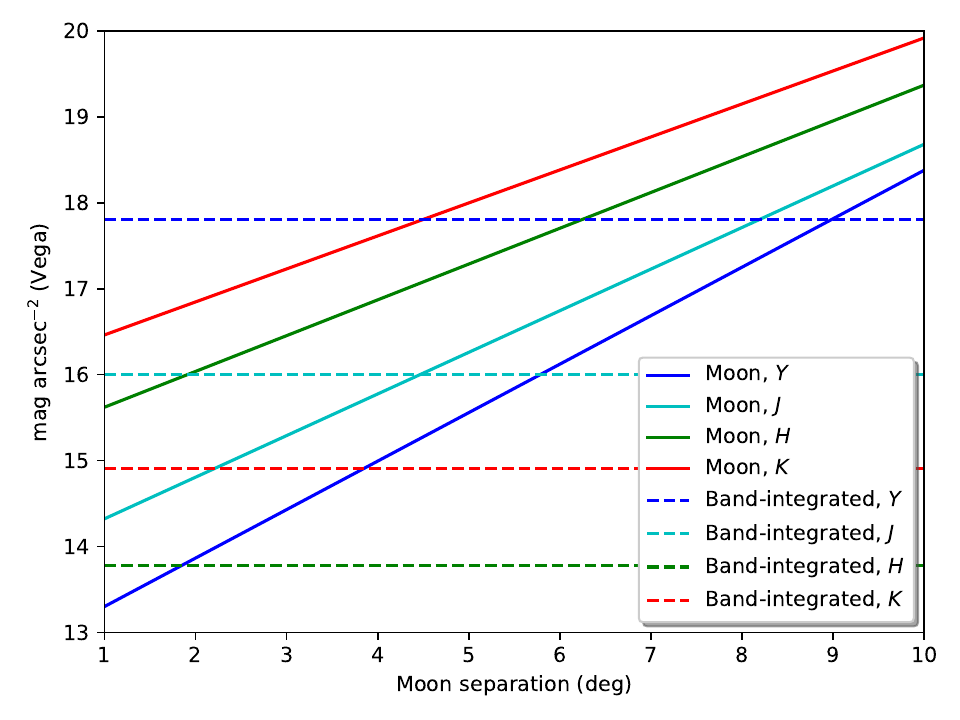}
\caption{Moon contribution as a function of the distance to the fiber expressed in mag arcsec$^{-2}$ in comparison of representative broad-band background ($JHK$ from \citealt{Srinath_swimming_2014}, $Y$ from \citealt{High_sky_2010}).The linear fits for the Moon background as a function of angular separation are listed in Table~\ref{tab:moonbgnd}.}
\label{fig:raw_bgnd3}
\end{figure}

\begin{table}[!h]
    \centering
    \begin{tabular}{|c|c|c|}
    \hline
        Band & Intercept & Slope (degree$^{-1}$) \\ 
        \hline\hline
        $Y$ & 12.74 & 0.56\\
        $J$ & 13.84 & 0.48\\
        $H$ & 15.20 & 0.42\\
        $K$ & 16.08 & 0.38\\
            \hline
    \end{tabular}
    \caption{Measured background as a function of angular separation from the Moon on our dataset for the four NIR standard bandpasses. Magnitudes are given in the Vega system and expressed if units of mag arcsec$^{-2}$.}
    \label{tab:moonbgnd}
\end{table}

\section{Results} \label{sec:results}

\normalsize{

\subsection{Spectra Analysis} \label{sec:spec_analysis}

The spectra medians at Events 1 and 2 were dominated by low-level persistence from daily calibrations. Although these effects of persistence were well handled by our flux model, the global temporal trends cannot be used to determine the continuum part of the sky spectrum. Figure~\ref{fig:raw_bgnd} shows the median background in $YJHK$ bandpasses on a sample night within Event~1; the multi-hour decay of persistence from the evening calibration is the dominant contribution for the first half of the night and persists at a low level through the entire night. The OH lines barely affected the median flux, meaning we probed other effects. The inter-line sky background was extremely faint, on the order of a millionth of our flux units, agreeing with other studies \citep{2015_oliva}. This interline background is small enough as to be negligible when compared with the band-integrated flux of all emission lines. In addition, the airmass was centered at $1.123\pm0.328$ for observations across the 3.5\,years, while the airmass was 1 for every observation in Events 1 and 2.

\subsection{Moon contribution to the near-infrared sky background}

The near-infrared sky background is dominated by OH lines, and the continuum contribution from the Moon is generally considered negligible and little attention has been paid to its contribution outside of the optical domain. From our spectroscopic time series, we extracted a median $Y$, $J$, $H$, and $K$ background level, which is relatively insensitive to the presence of OH lines, these covering only a few \% of each bandpass. The median flux level on the 3rd night is shown in Figure \ref{fig:raw_bgnd}, with the combination of a slow decay over the night, which is due to the daily calibration, most notably the flat field frames. This persistence is a well-known feature of Hawaii arrays \citep{2018_artigau}. In addition to the persistence decay, there is an increase in the continuum level at the end of nights that is due to the serendipitous passing of the Moon within $\sim1^\circ$ of SPIRou's fiber. This allows for the first measurement of the background contribution of the Moon at infrared wavelengths on Maunakea.

Observations affected by the Moon were obtained on 2019-12-14, 2019-12-15 and 2019-12-16. The full Moon happened on 2019-12-12 at 00:12 and observations were obtained 2 to 4 days after the full Moon with an illumination ranging from 95\% to 80\%.

For each of the 3 nights for which we see the Moon contribution, we computed the mean distance to the Moon for the fiber and subtracted a linear fit of the persistence contribution for angles ranging from 50 to $20^\circ$. The background level was then flux-calibrated by computing the median flux derived from the observations of telluric standard stars of known $YJHK$ magnitudes. Finally, fluxes were expressed in magnitudes (Vega system) per square arcsecond, considering SPIRou's $1.29\arcsec$ fiber diameter. There is a clear decay of the background with distance to the Moon (see Figure \ref{fig:raw_bgnd2}), and our observations can track its continuum contribution to about $6^\circ$ in all bands.

The background level derived from these observations is shown in Figure \ref{fig:raw_bgnd3}. The Moon contribution is smaller than the band-integrated background in $H$ and $K$, but it dominates within $\sim4^\circ$ in $J$ and $\sim8^\circ$ in $Y$. The effective magnitude of the continuum contribution by the Moon is an important input for survey strategies for multiplex spectrographs. If one uses a PCA reconstruction of the sky spectrum and attempts to observe targets at $YJHK$ mags fainter than $\sim$15 and attempts a sky subtraction better than 1\%, then the Moon contribution must be accounted for at Moon separation distances of at least 10$^\circ$. However, these angular distances can change if one considers the variation of atmospheric aerosol since small scattering angles can contaminate the near-IR \citep{2013A&A...549A...8B}.

\subsection{Line Fitting Analysis} \label{sec:flux_analysis}

In this subsection, we summarize the trends from our optimal parameters and metrics during model fitting. All parameters and metrics were stationary with scatter introduced by poor fitting unless stated otherwise.

\subsubsection{Local Median and Median Absolute Deviation}\label{sec:local_median}

The local medians for the lines all followed the same trends as the global medians. The reddest and bluest lines captured the systemic effects described earlier in Section \ref{sec:dataset}. These effects positively biased the local medians of our reddest lines, and negatively biased the local medians of our bluest lines.
The local $MAD$s generally increased in the bluer lines since they contained more noise than the redder lines.
The local $MAD$s sharply decreased on the last two days of Event 1 for a period of an hour, similar to the $K$-band decrease during that time.

\subsubsection{Total Flux Contribution and Amplitude ($A_i, \hat{F}_{flux}(\lambda)$)}\label{sec:amp}

The flux from each OH line, i.e. total flux contributions, generally followed a consistent, smooth, and continuous pattern, especially Events 1 and 2. The bluest lines often measured negative flux as a result of the noise. OH lines in the $H$ band (middle of the spectrum with high SNR) recorded some of the highest fluxes in both Events 1 and 2, which was expected. Since OH had high Einstein A-coefficients and relatively high populations in the low vibrational upper states, the  $\delta$v=2 bands (3-1), (4-2), and (5-3) in the $H$ band was the strongest \citep{2000_rousselot,oli2013, oli2015}}. 
Since the amplitude was bounded by $\pm$ one order of magnitude from the line's initial amplitude, all values were bounded by $\pm 1000$. Since the bluer lines were noisier, the fitted amplitudes were more likely to be negative. Similar to the flux, amplitudes followed the same temporal patterns. Noisy fits and spurious amplitudes from fitting to the background for singlets added scatter throughout the amplitude space.

\subsubsection{Line Center and Width ($\mu_i, \sigma_i$)} \label{sec:mu_sigma}

Since the line center parameter was bounded by $\pm1.5\times10^{-4}\mu$m, there were globally no trends during the entire observing period, including Events 1 and 2, as the spectral resolving power was too low to accurately measure that variation. Each $\mu_i$ was on average within $5\times10^{-5}{\mu}m$ of the recorded lines in the literature.

The line widths were centered on $10^{-5}{\mu}m$, and generally increased in the redder lines. Since the instrument resolution only varies $\sim5\%$, most of the increase was caused by SPIRou's constant resolution throughout the wavebands. The upper bound, we imposed on the fits, of $0.001\mathring{A}$ was rarely reached by a relatively small number of fits. Similar to amplitude, noisy fits and spurious parameters from fitting to the background for singlets added scatter throughout each parameter space.

\subsubsection{Errors, Correlations, and Statistical Significance ($\eta$, $MSE_F$, $R_{Pearson}$, $R_{Spearman}$, p-value)}\label{sec:error_corr}

The relative error was centered $\approx1\%$ with $\approx86\%$ of the errors being $<5\%$, suggesting agreement between the observations and fits. 

The Pearson correlation was centered $\approx0.88$ with $\approx86\%$ of the correlations being $>0.5$. The redder lines generally had higher correlations than the bluer lines. The Spearman correlation between the fit and data generally increased for the redder lines with $\approx58\%$ of the fits' correlations being $>0.5$. $\approx86\%$ of the fits' p-values where statistically significant ($<0.05$), and was centered $\approx2\times10^{-6}$ ($>4.5\sigma$).

\subsection{Resolved Hydroxyl Doublets}

Due to the high spectral resolution of SPIRou and the hundreds of spectral fits per observation, we used a data-driven approach to observe doublets that may have been previously observed as singlets in Maunakea's atmosphere. A doublet (see bottom-left of Figure \ref{fig:model_fits} for an example) is caused by Lambda doubling, and splits a singlet (see top-left of Figure \ref{fig:model_fits} for an example) into two separate components with similar flux symmetric about the singlet line center \citep{1998JQSRT..60..665R, 2022JQSRT.27707949G}.
We note that due to Lambda doubling, all OH emission lines are doublets. However, the separation between both components can vary a lot depending on the line parameters and hence the numbers of detected doublets depend on the spectral resolution of the spectrograph used to observe them. The most accurate OH line positions can currently be found in the \href{https://hitran.org/}{HITRAN} database (\citet{HIT2022}).

We estimated the probability distribution of each line center (i.e. $\mu_1$ and $\mu_2$) using a Gaussian kernel density estimator (KDE) with a bandwidth of 0.05. The two global maxima from the estimator were used as the estimates for $\mu_1$ and $\mu_2$ for each OH line. Next, we found the sum of the difference between the estimated line centers and the line center from \cite{2000_rousselot} (i.e. $\mu_1 - \mu_{measured} + \mu_2 - \mu_{measured}$) as a symmetry metric. If the estimated line centers are symmetric about the measured OH line, then this sum is 0. Then, we found the ratio between the KDE evaluated at each estimated line center (i.e. KDE($\mu_1$) / KDE($\mu_2$)) as a similarity metric. If the difference between the estimated line centers was $\pm0.05\mathring{A}$ and the ratio between the fluxes associated with each component was between 0.78 and 1.3 (i.e. less than 30\% difference), then the line was classified as a resolved doublet. The identified doublets are given in Table \ref{tab:doublet_all}.
We note that that doublets with wide component separation tend to have clear differences in the intensities of both components \citep{noll23, 2020ACP....20.5269N}. \citet{noll23} find ratios outside the range from 0.78 to 1.3.

\subsection{Short-term Variability: GPR Analysis\label{section:gpr}}

\begin{figure}[!htbp]
\centering
\includegraphics[width=0.8\linewidth]{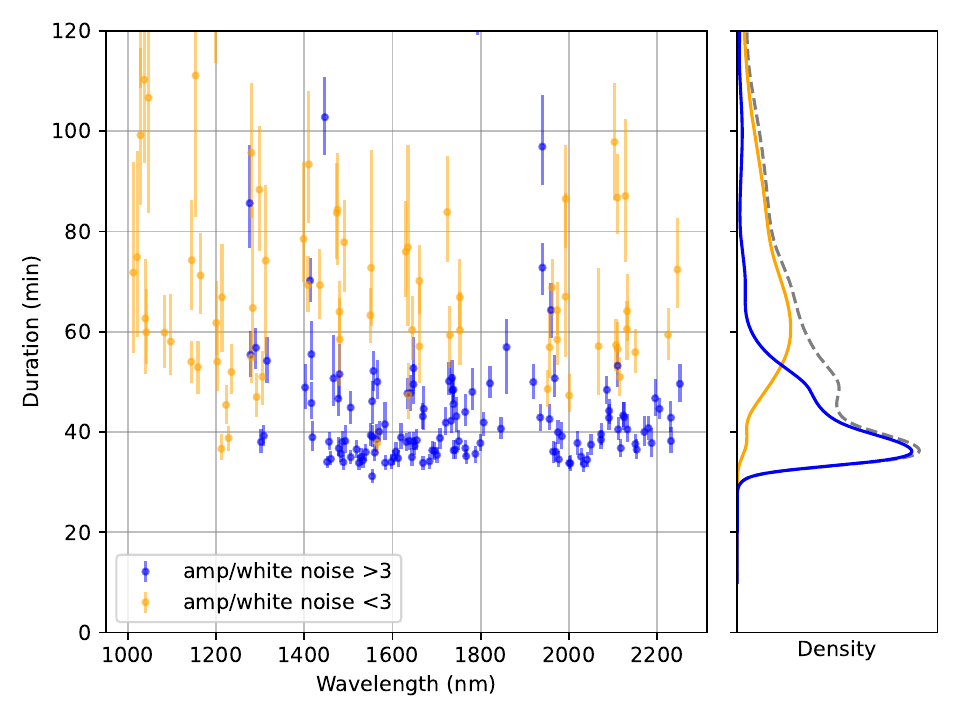}
\caption{Correlation length of the GPR fit for OH line time series. There is a correlation between the ratio of the GPR amplitude over the white noise term ($\sigma_{GPR}/\sigma_{\rm{wn}}$) and the correlation length. The inclusion of a large white noise term masks short-term variability in the time series. Overall, there is a consistent 'floor' in the distribution of correlation timescales around $\sim30$\,min, with a median timescale for high-SNR lines of 39\,min. }
\label{fig:distrib_lengths}
\end{figure}

To investigate the timescale of OH intensity variations in our data, we used a simple GPR model to quantify the time correlation of the brighter lines in our times series. We focused on the 6 nights with near-continuous monitoring for each line's time series. The GPR regression gives a correlation length of $\sim$30\,min with a relatively small white noise contribution compared to the GPR amplitude. The distribution of correlation lengths is shown in Figure \ref{fig:distrib_lengths} and has a median value of 39\,min for lines having a large amplitude relative to the white noise term ($>3$). Longer timescales are generally derived for lines that have a lower SNR as the white noise contribution masks short-term variability.

The GPR framework is intended to capture the shortest timescales for emission line variability. It is these timescales that matter in establishing sky-subtraction strategies. Beyond the observed timescales of tens of minutes, there is a night-long variation of the mean sky level \citep[e.g.,][]{2013AJ....145...51T}, night-to-night variations and one also expects seasonal with varying solar illumination. While relevant in the description of sky background physics, these timescales do not come into play in the design of spectroscopic observation sequences. 

Using the GPR fit of the time series, we can determine the fractional error that one would have in the estimate of the sky as a function of the time delay between the sky measurement and the science observation. This is done by determining the RMS of the ratio of two points at an increasingly large time difference. We performed this analysis for all bright OH lines and determined the timescale after which one would incur a 1\%, 2\%, and 5\% fractional error in the sky correction. The median timescale for a 1\% is very small ($\sim$2\,minutes), shorter than the spectra-to-spectra time in our sequences. Even a 5\% fractional error would require a sky sampling time of $\sim$10\,minutes. In practice, most deep near-infrared spectroscopic observations use a sky-subtraction with the ABBA pattern. The sky is observed before and after the science exposure (or a nodding is applied in slit-spectroscopy between observations) and one effectively accounts for the first-order time-variability of the sky spectrum. One is therefore only left with higher-order time variations. From the GPR fit of each line, we determined the timescale corresponding to a 1\%, 2\% and 5\% sky-subtraction error in an ABBA scenario. The timescales are much more manageable in real-life observing strategies, with a $\sim$10\,min delay between the first and last sky observation required for a 1\% residual amplitude. 

\begin{figure*}[!htbp]
\centering
\includegraphics[width=0.98\linewidth]{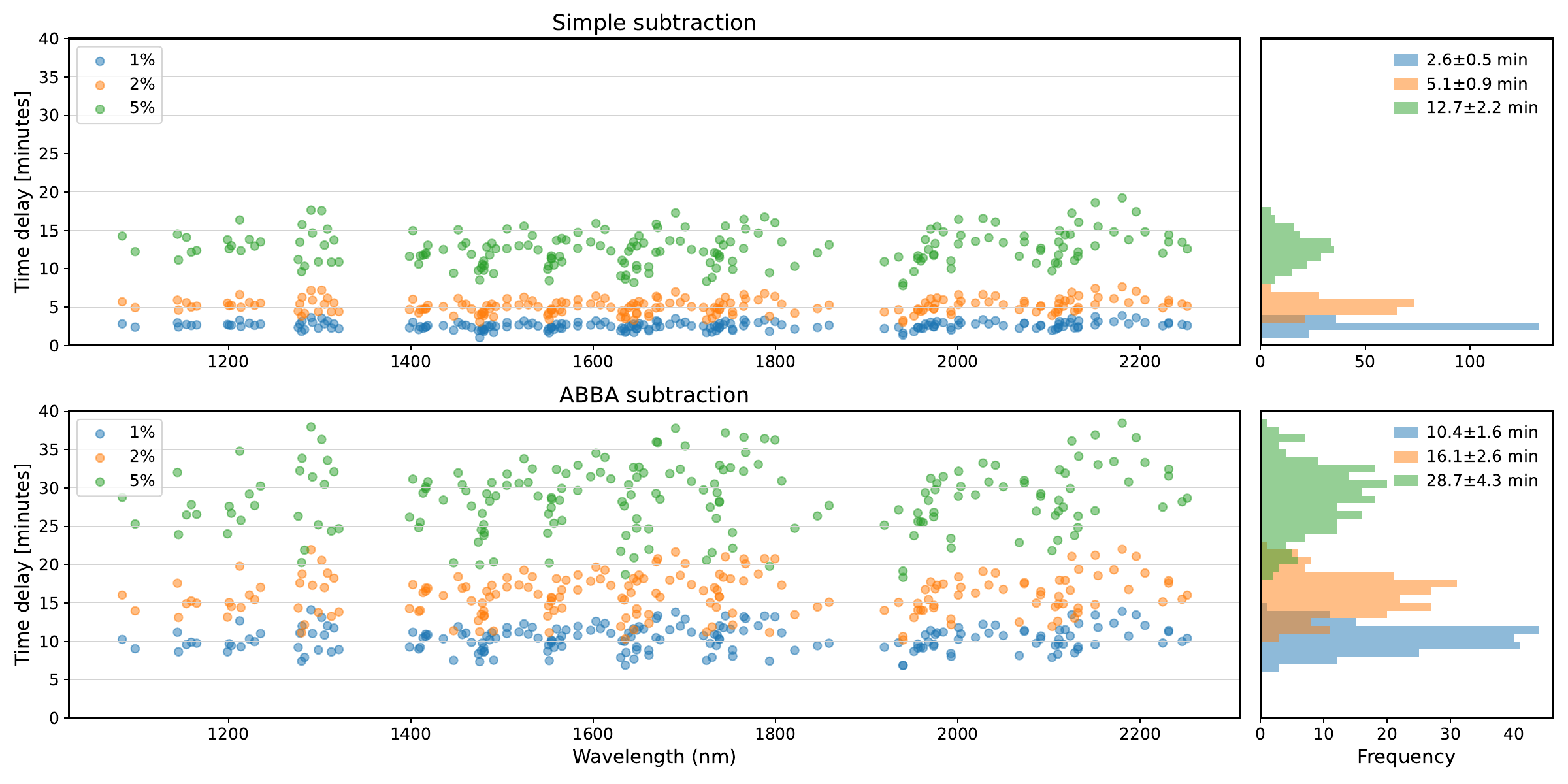}
\caption{Timescale for an RMS error in sky subtraction of 1\%, 2\% and 5\% fractional accuracy. The top panels show the case of a naive sky subtraction while the bottom panels show the case with an ABBA sky subtraction that handles sky line changes that are linear with time. An accuracy of 1\% can be obtained with a sky sampling time of $\sim$10\,min with an ABBA subtraction while a $\sim$2.6\,min sampling would be required for a simple (i.e., not accounting for time variation) subtraction.}
\label{fig:etarget' psilon}
\end{figure*}

\clearpage

\subsection{Long-term Variability: LSP Analysis} \label{sec:lsp_analysis}

\begin{figure}[t]
\centering
\includegraphics[scale=0.14]{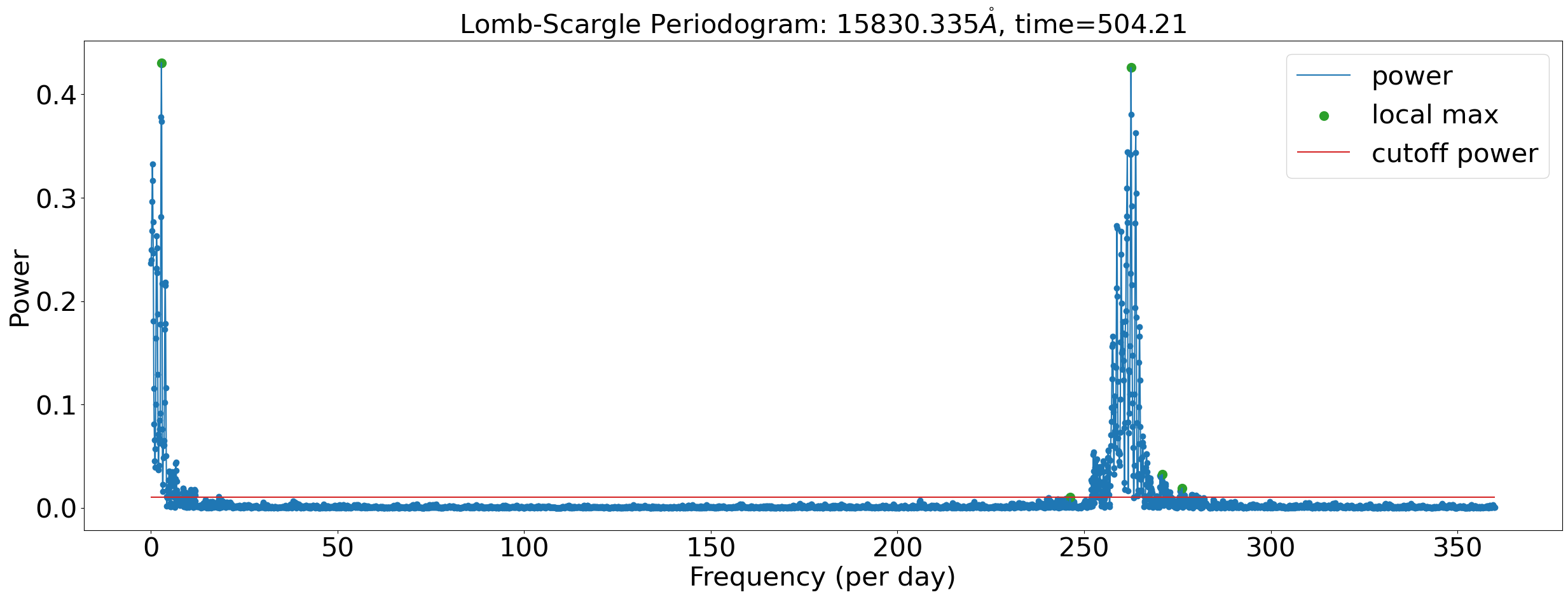}
\centering
\includegraphics[scale=0.14]{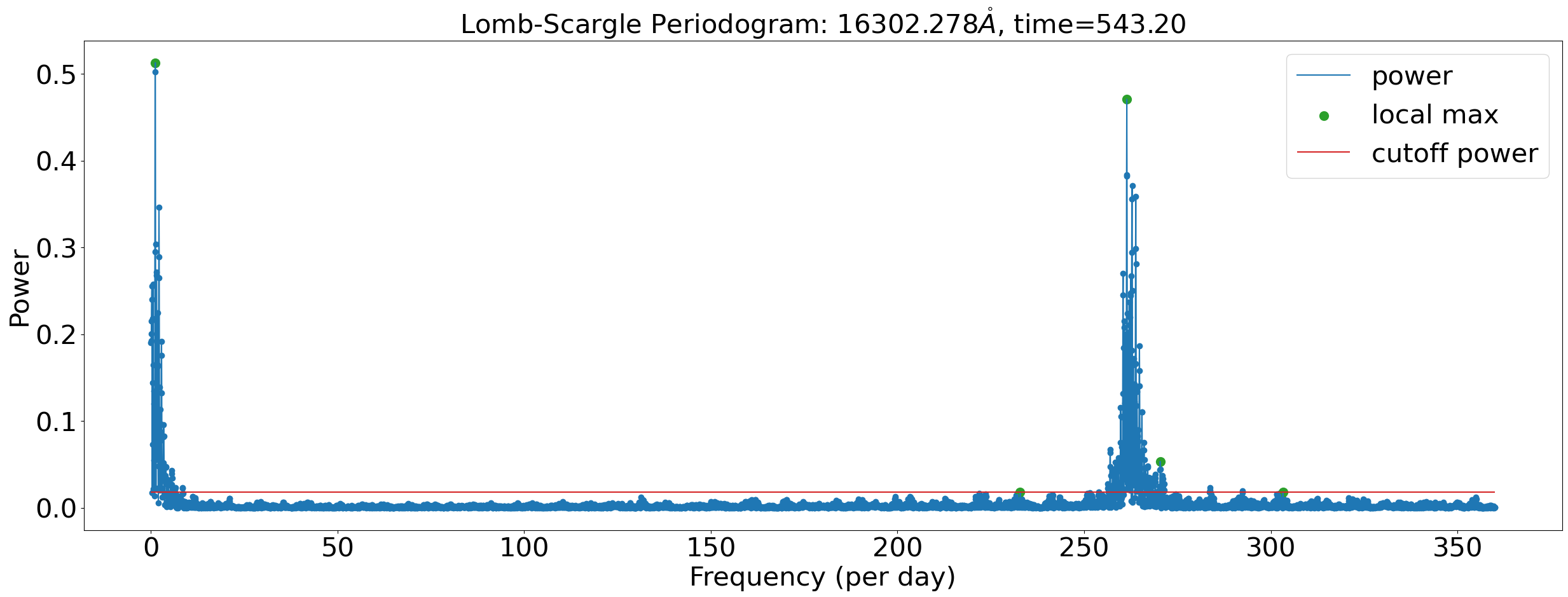}
\centering
\includegraphics[scale=0.14]{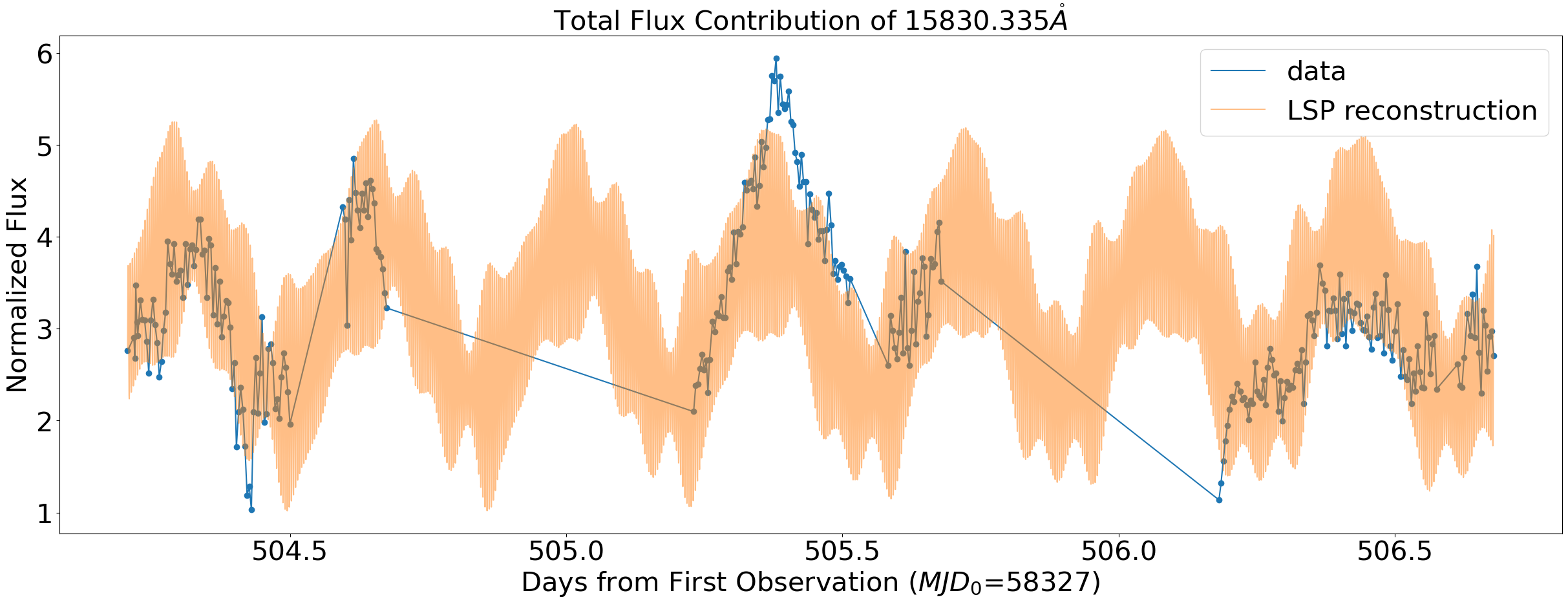}
\centering
\includegraphics[scale=0.14]{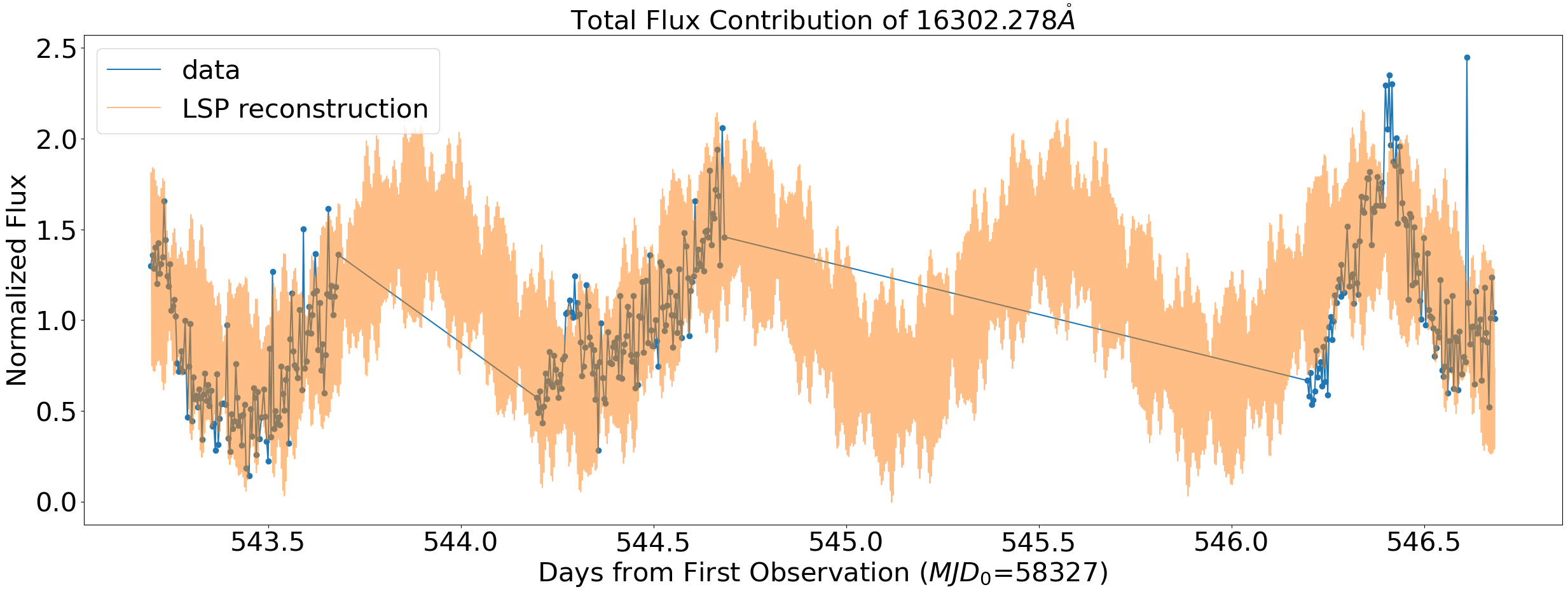}
\caption{Examples of periodograms (top) and reconstructions (bottom) from Event 1 (left) and 2 (right). The red line indicates the cutoff power to be considered a relative maximum. The periodograms were concentrated at periods of 1 day and 5.5 minutes. Because 5.5 minutes is the typical cadence for the sky observations used here, this is not a measurement of the NIR sky variability but a result of our sampling rate.}\label{fig:lsp}
\end{figure}

\begin{figure}[t]
\centering
\includegraphics[scale=0.4]{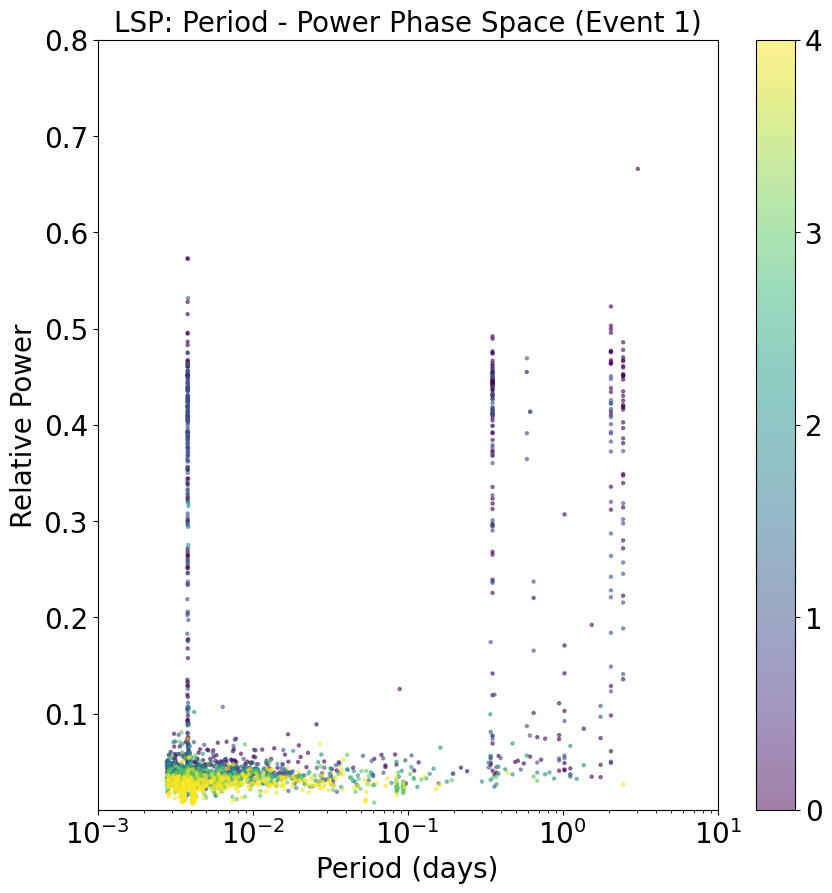}
\centering
\includegraphics[scale=0.4]{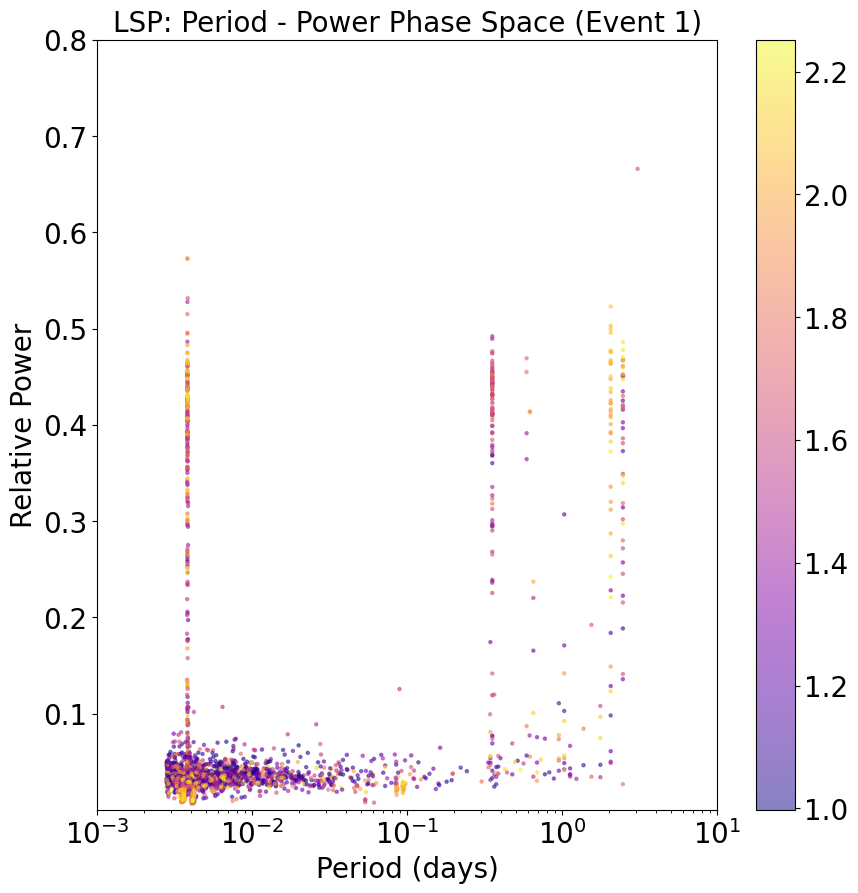}
\caption{Lomb-Scargle Period-Power Phase Diagrams for Event 1 colored by period order (left) and wavelength (right). The first-period order was centered around 1/3 of a day (8 hours) and 2 days while the second and third-period orders were centered around 5.5 minutes (i.e. spurious peak at the sampling rate). A majority of the emission lines falling under the same periods, in general, indicated that most OH lines vary homogeneously. The large peak at log(p)=-2.4 corresponds to exactly the sampling rate of the data, which is 5.5 min. We note that with a periodogram search there is a point where the folding interleaves two nights such that one night is at a phase of 0 and the other of at a phase of 0.5. This leads to a false periodicity very close to your sampling rate. We emphasize that it's not possible with the dataset to properly constrain timescales beyond $\sim$0.3 days which suggests that if a survey seeks to optimize sky-subtraction strategies, sky data on smaller time-scales finer time resolution sky data are needed.}

\label{fig:lsp_event1}
\end{figure}

\begin{figure}[t]
\centering
\includegraphics[scale=0.4]{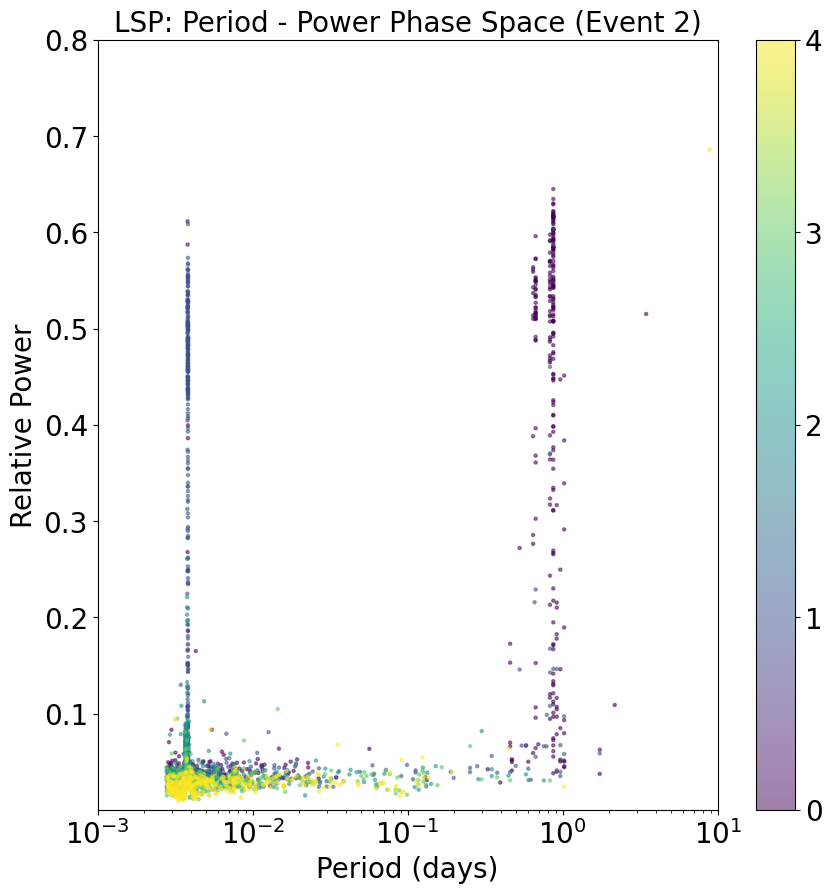}
\centering
\includegraphics[scale=0.4]{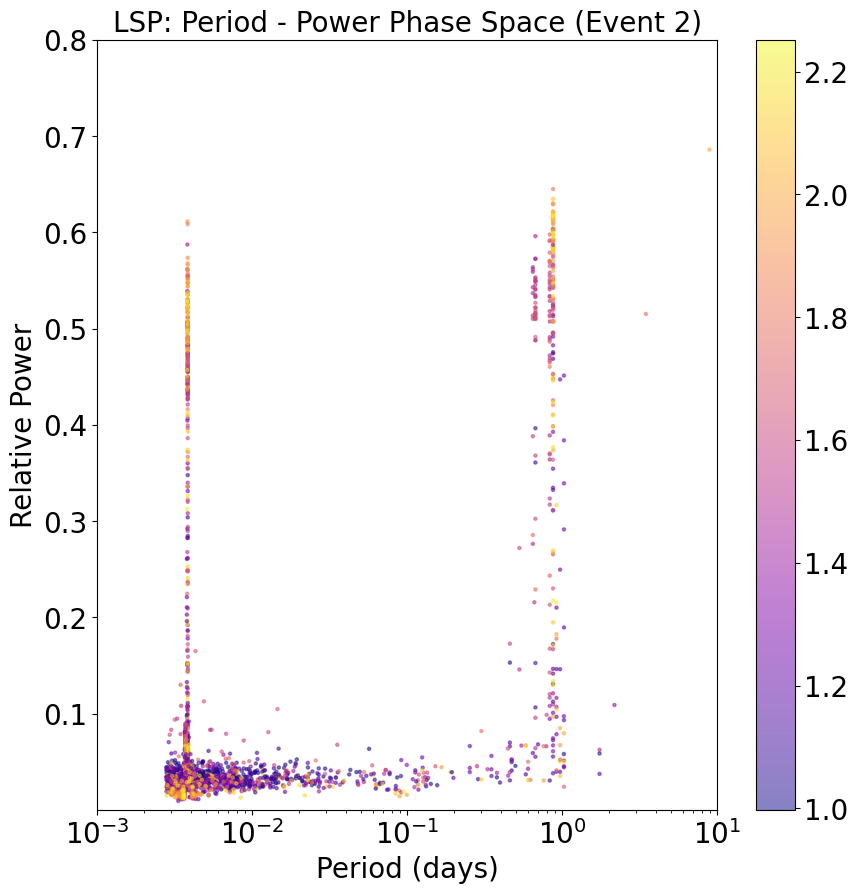}
\caption{Lomb-Scargle Period-Power Phase Diagrams for Event 2 colored by period order (left) and wavelength (right). The first period order was centered around a day while the rest of the periods were relatively small periods which are due to our sampling because we obtained observations every $\sim$5 minuutes. Similar to Event 1, the redder wavelengths also had the largest powers. \label{fig:lsp_event2}}
\end{figure}

Figure \ref{fig:lsp} illustrates a few reconstructions using the most dominant periods from the periodograms. The maximum frequency searched was 360 cycles/day so in turn the minimum period was 1/360=0.003 days or 4 minutes. The periods in days for Event 1, from most to least dominant by power (i.e. period order), were $0.60\pm0.86$, $0.41\pm0.72$, $0.08\pm0.21$, $0.03\pm0.07$, and $0.09\pm1.11$, respectively. We did not include periods $<6$ minutes since those were caused by the mean period of observation (5.5 minutes). In Event 1, the OH sky lines varied on the order of half a day. This can be explained with a folding of consecutive nights. A signal at a sub-hour timescale will not survive from one night to another in this context.  The sky was observed during Events 1 and 2 for approximately half a day, and a cycle was completed within that time. Figure \ref{fig:lsp_event1} visualizes the period-power phase space for Event 1. Most of the variability was captured within the first order, while the next four were kept in an attempt to increase fit quality and metrics. Most of the dominant periods were centered around one-third of a day (8 hours) and two days. The second and third order (second and third most dominant) periods were centered around 5.5 minutes, while the rest were scattered along short periods ($<0.1$ days). The distribution of periods expressed the infrared sky's complexity. The periods spread evenly in general across all wavelengths with the $H$ band hosting the best period fits as explained in Section \ref{sec:spec_analysis}.

The periods in days for Event 2, from most to least dominant by power, were $0.61\pm0.42$, $0.18\pm0.33$, $0.04\pm0.11$, $0.03\pm0.08$, and $0.36\pm2.66$, respectively. Similarly to Event 1, all of the least dominant frequencies were less than 0.1 days. In addition, the period's spread as a function of wavelength was also similar. However, the OH lines varied primarily in order of a day instead of split between two periodic modes as in Event 1. The powers were higher on average and varied less in Event 2. Figure \ref{fig:lsp_event2} visualizes the period-power phase space for Event 2.

The reconstructions for Event 1 generally had smaller $MSE$s than Event 2 ($19.554\pm54.772$ vs $32.508\pm114.660$). However, Event 1 had a slightly lower $R_{pearson}$ than Event 2 ($0.510\pm0.189$ vs $0.517\pm0.211$). These two metrics should be inversely proportional to each other so this result was unexpected. Event 1 better reconstructed the total flux contributions, but Event 2 better captured the overall trends. The most dominant periods, oscillation period and observation cadence, were responsible for the overall trends while the least dominant periods were responsible for the smaller details. Overall the analysis is limited by the impact of our sampling rate and a lack data at intermediate scales. Although we cannot properly constrain timescales beyond $\sim0.3$ days, the power spectra suggested substantial variability on the order of a day. Further analysis may prove more valuable when combined with additional atmospheric conditions data, and cover a wider range of timescales.

\section{Discussion} \label{sec:discuss}

\subsection{Further Improvements} \label{sec:improve}

Our aim was to characterize the sky emission by fitting Gaussian profiles to and investigating the possibility of identifying timescales for sky variations that can guide observing strategies in the NIR. However, we had only 1075 observations of which 720 were used for the full analysis because they were high time-resolution observations. As a result, we only made 481 OH lines $\times$ 720 spectra $= 346,320$ individual OH line measurements. Obtaining more data would position us to constrain our results to a better degree. Our flux model was generally sufficient, but two Gaussians after a low-pass filter may be too simple of a model to capture the rest of the emission lines' features. Expanding the flux model to more complex profiles and continuum fitting may be necessary to capture the true flux of the OH lines. In addition, the periodograms were able to capture a good portion of the variability, but the model did not fit the data well. Other time series analysis algorithms may lead to better fitting results. 

\subsection{Guidelines for infrared spectroscopic observations \label{guidelines}}

The level of sky residuals that one will have in a science observation with an on/off approach to sky measurement will depend on the timescale for sky nodding and the brightness of the science target (expressing sky as a fractional error). The longer the delay between on and off-sky observations, the worse the sky subtraction. By adjusting a GPR on the time-varying intensity of OH lines, we determined that all bright lines have a variation timescale of 30\,min and larger, which in turn provide a single relation regarding the quality of subtraction we can obtain considering a given delay to measuring the sky. Allowing a 1\% error in fractional line intensity measurement, observing sky every 3\,min is necessary. This assumes that the sky subtraction is simplistic without a linear interpolation: one obtains a science observation, a sky frame and subtracts the two. If one allows for an interleaved sky-science sequence (such as in the ABBA pattern), the timescale for a 1\% error increases as the linear change of sky flux with time is accounted for, leaving only higher-order terms.

\section{Conclusions \label{sec:conclude}}

Developing precise and data-driven sky subtraction algorithms is a critical task for the next decade of ground-based astronomical calibration and observation. Hydroxyl line sky emission in the infrared varies as a function of time and its complexity is not well understood.

We present a a unique set of 1075, Maunakea skies spectra that have high spectral resolution which includes six nights of dedicated sky observations, taken every five minutes. To further leverage the use of this data set  we make it available in Zenodo, together with the line measurements of hydroxyl lines.

We also present the first spectroscopic estimate for the Moon's contribution at infrared backgrounds on Maunakea. This estimate is the result of finding an increase in the continuum level at the end of nights due to the serendipitous passing of the Moon within $\sim1^\circ$ of SPIRou's fiber. Our approach and results are complementary to modeling work by \citet{kstone2005} and \citet{2013_jones}. \citet{kstone2005} investigate how the lunar phase affects the albedo and hence the expected NIR background to 2.38$\mu$m. \citet{2013_jones} provide the model for the Moon's contribution for visible to NIR observations, and is an important part of the ESO sky model which helped revolutionize the way ESO facilities perform sky subtraction. See also \citep{moh14, smette2015, kau2015, 2019A&A...624A..39J, noll24}.
 
We test several analysis tools to characterize the variability of the Maunakea NIR night sky. From GPR, we measured a $\sim30$ min correlation length, and determined the timescale to achieve 1\% fractional error for an ABBA scenario was $\sim10$ min. An LSP analysis is limited by the sampling rate but detects power on multiple scales from minutes to hours. This analysis is limited by our sampling rate and simultaneous detailed weather monitoring but provides a significant step toward developing a Maunakea sky model similar to the very successful $ESO~SkyCalc$ set of tools which have been shown to improve sky subtraction for NIR spectra \citep{pat2008, 2014_noll}. 

\section{Acknowledgments}
This work is based on observations made by the Canada-France-Hawaii Telescope. We acknowledge Takahiro Morishita for early discussions on the applicability of this work to several Maunakea facilities. We acknowledge Néstor Espinoza for advice on the implementation of Lomb-Scargle Periodograms and Gaussian processes. 

\noindent{MN is supported in part by NASA under award No. 80NSSC22K0821. FD \& AP acknowledge support for this work from an STScI DDRF grant. The authors wish to recognize and acknowledge the very significant cultural role and reverence that the summit of Maunakea has always had within the Native Hawaiian community. We are most fortunate to have had the opportunity to conduct observations from Maunakea. As we grapple with climate change, fragile island environments like those of Hawai\textquoteleft{i} are in peril. Finding sustainable environmental approaches is urgent for all of us. Ancient Hawaiians understood how to create balance and harmony between people and the Earth; knowledge was passed down through the concept of aloha \textquoteleft{aina}  \citep[e.g.,][]{aina}. We recognize the wisdom of generations of Indigenous Hawaiians who shaped Hawai\textquoteleft{i} and hope to learn from them.}

\vspace{5mm}
\facilities{MKO, CFHT(SPIRou)}

\software{Astropy \citep{2018_astropy},  
          Emcee \citep{2013_foreman},
          George \citep{2015ITPAM..38..252A},
          Matplotlib \citep{2007_plot},
          Numpy \citep{2020_numpy}, 
          Pandas \citep{2010_pandas},
          Scipy \citep{2020SciPy-NMeth},
          }

\appendix\label{sec:appendix}
\counterwithin{table}{section}
\counterwithin{figure}{section}

\section{Data Availability}

\begin{figure}[t]
\centering
\includegraphics[scale=0.5]{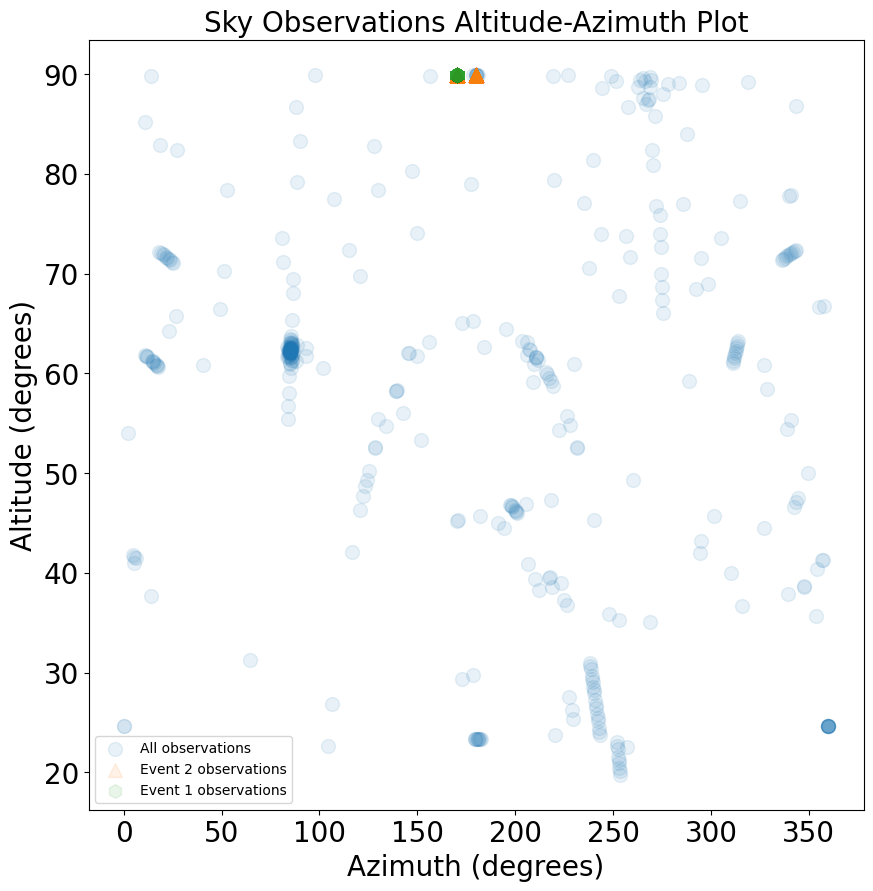}
\caption{The altitude-azimuth coordinates for the sky observations. An altitude of $90^\circ$ points directly overhead, and an azimuth of $0^\circ, 90^\circ, 180^\circ,$ and $270^\circ$ points north, east, south, and west, respectively. Darker data points correspond to more observations and Events 1 and 2 refer to the two three-night runs in December 2019 and January 2020 when we acquired sky spectra every five minutes. These sky observations provide a unique set of high cadence and long-duration data with which to investigate the variability and correlations of NIR sky lines.} \label{fig:obs_coords}
\end{figure}

We made the data used for this study available through Zenodo \dataset[doi:10.5281/zenodo.13363061]{https://doi.org/10.5281/zenodo.13363061}, which is a popular data repository used in many domains, including astronomy. The repository contains a zip file titled \texttt{spirou\_sky.tar.gz}, which contains the 1075 sky observations in the FITS file format. Figure \ref{fig:obs_coords} summarizes the celestial coordinates for the sky observations. The repository also contains a text file titled \texttt{transition\_doublet\_table.txt}, which is a digital version of Table \ref{tab:doublet_all}. Along with our code, the OH line measurements from \cite{2000_rousselot}, a median stacked sky, and our various measurements are made available through \href{https://github.com/FDauphin/spirou-sky-subtraction}{GitHub}.

\clearpage
\begin{deluxetable*}{ccccc}
\tabletypesize{\footnotesize}
\tablecolumns{5}
\tablewidth{0pt}
\tablecaption{Doublet Candidates \label{tab:doublet_all}}
\tablehead{\colhead{Transition} & \colhead{$\mu 0$} & \colhead{$\mu 1$} & \colhead{$\mu 2$} & \colhead{Doublet} \\}
\startdata
$[9-5]R2e(0.5)$  & 9975.083 & 9974.239 & 9975.086 & No \\
$[3-0]P2f(4.5)$ & 10002.737 & 10002.770 & 10003.521 & No \\
$[9-5]Q2e(0.5)$  & 10014.086 & 10013.711 & 10014.011 & No \\
$[9-5]Q1e(1.5)$  & 10015.555 & 10015.144 & 10015.546 & No \\
$[3-0]P1e(5.5)$  & 10016.731 & 10016.884 & 10017.124 & No \\
$[9-5]Q2e(1.5)$  & 10024.084 & 10022.844 & 10023.955 & No \\
\ldots          & \ldots     & \ldots    & \ldots & \ldots \\
$[9-7]P1f(5.5)$ & 22313.650 & 22312.692 & 22313.665 & No \\
$[9-7]P2f(5.5)$ & 22460.264 & 22459.955 & 22460.645 & No \\
$[9-7]P1e(6.5)$  & 22516.727 & 22513.733 & 22516.748 & No \\
\enddata
{\tablecomments{The columns are as follows: column (1) gives the transition from \citep{oli2015, 2000_rousselot}, column (2) is the initial guess for the fits to the line and, columns (3) and (4) are best-fit line centers, and column (5) is a Yes/No flag noting if the line was identified as a doublet. Columns 2-4 are in Angstroms.
The complete table is available online at \href{https://github.com/FDauphin/spirou-sky-subtraction}{GitHub}, on Zenodo: \dataset[doi:10.5281/zenodo.13363061]{https://doi.org/10.5281/zenodo.13363061}, and as a machine-readable in the online Journal. }}
\end{deluxetable*}

\section{Total Flux Correlation Matrices}

\begin{figure}[t]
\centering
\includegraphics[scale=0.17]{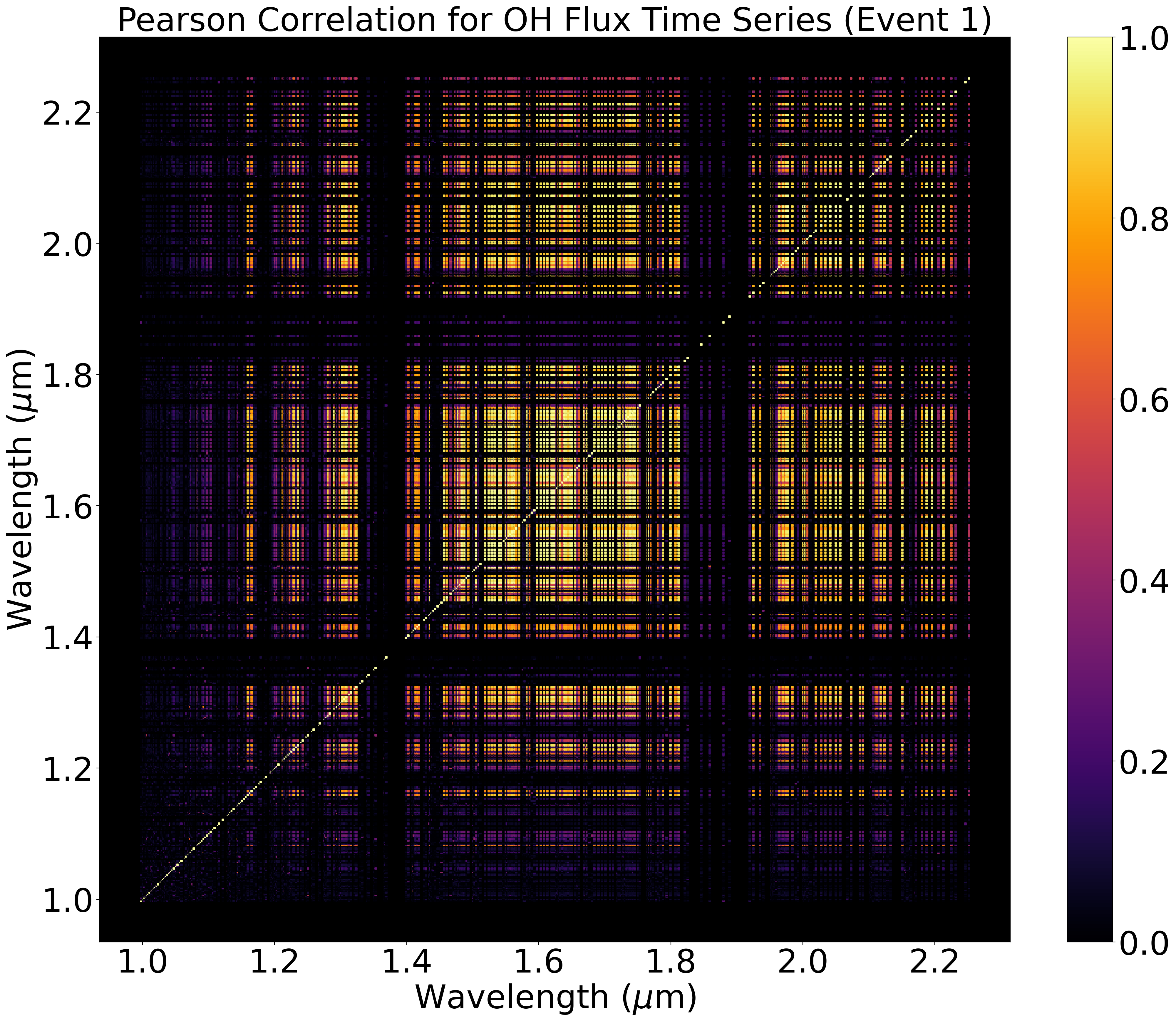}
\centering
\includegraphics[scale=0.17]{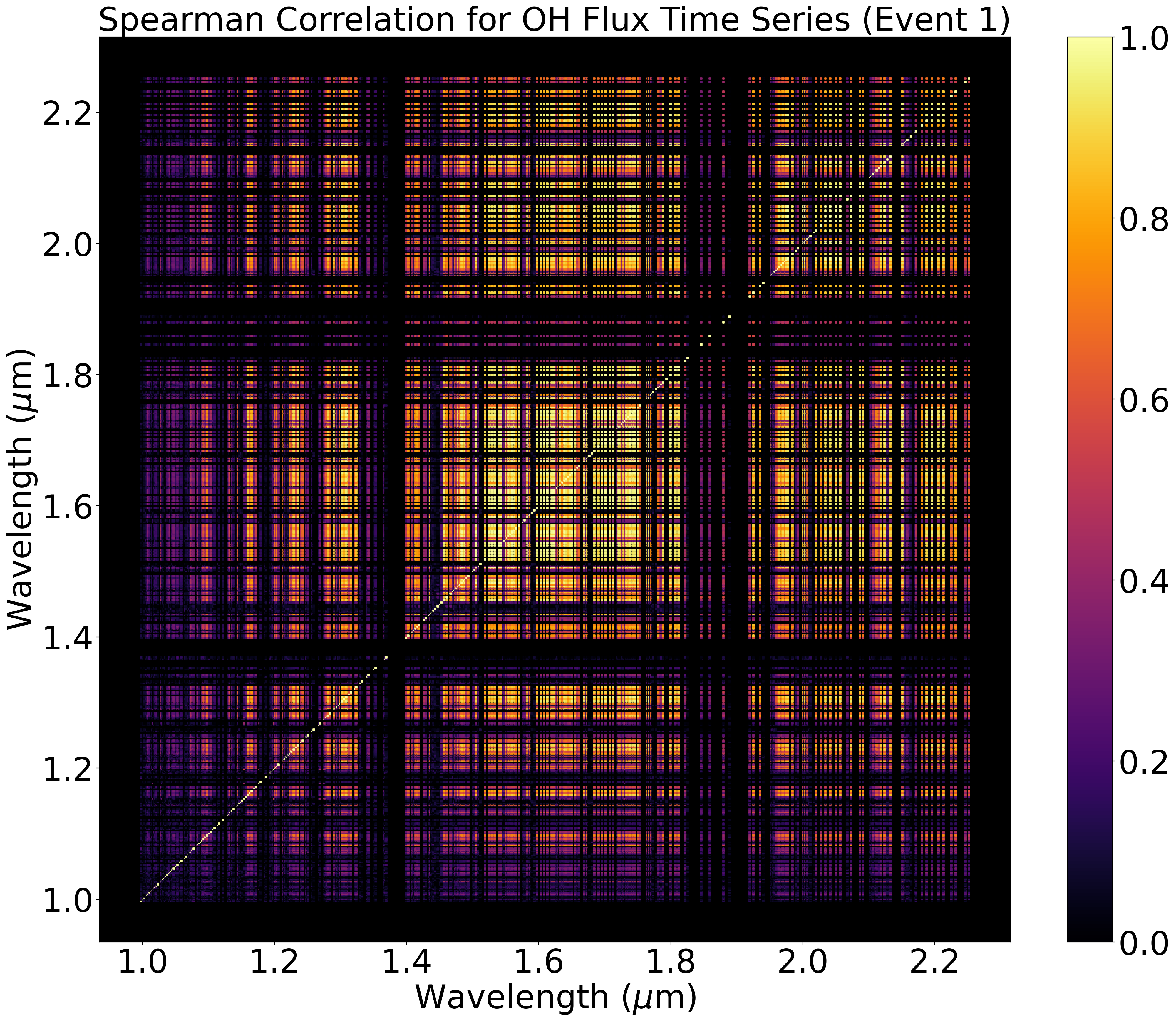}
\centering
\includegraphics[scale=0.17]{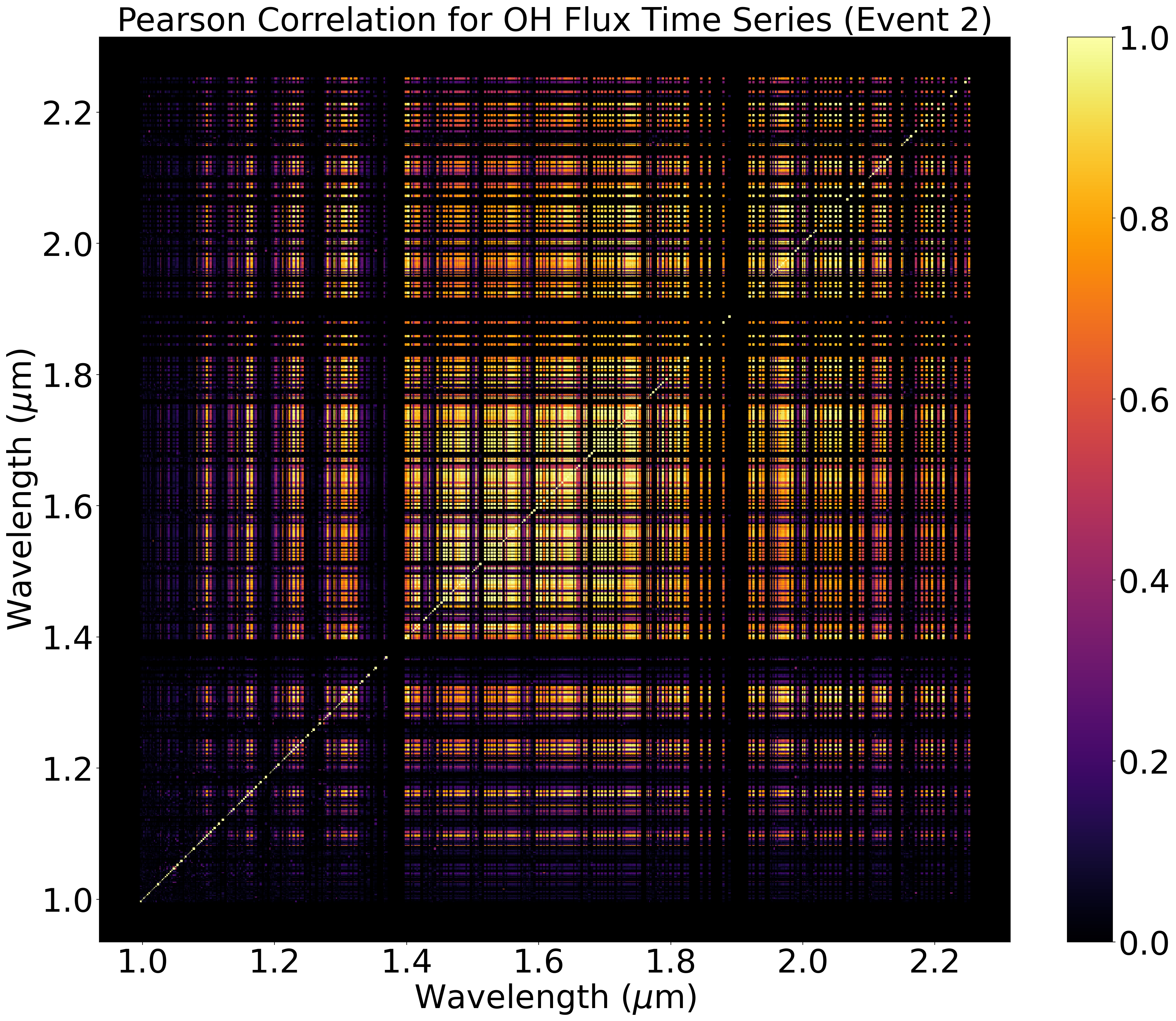}
\centering
\includegraphics[scale=0.17]{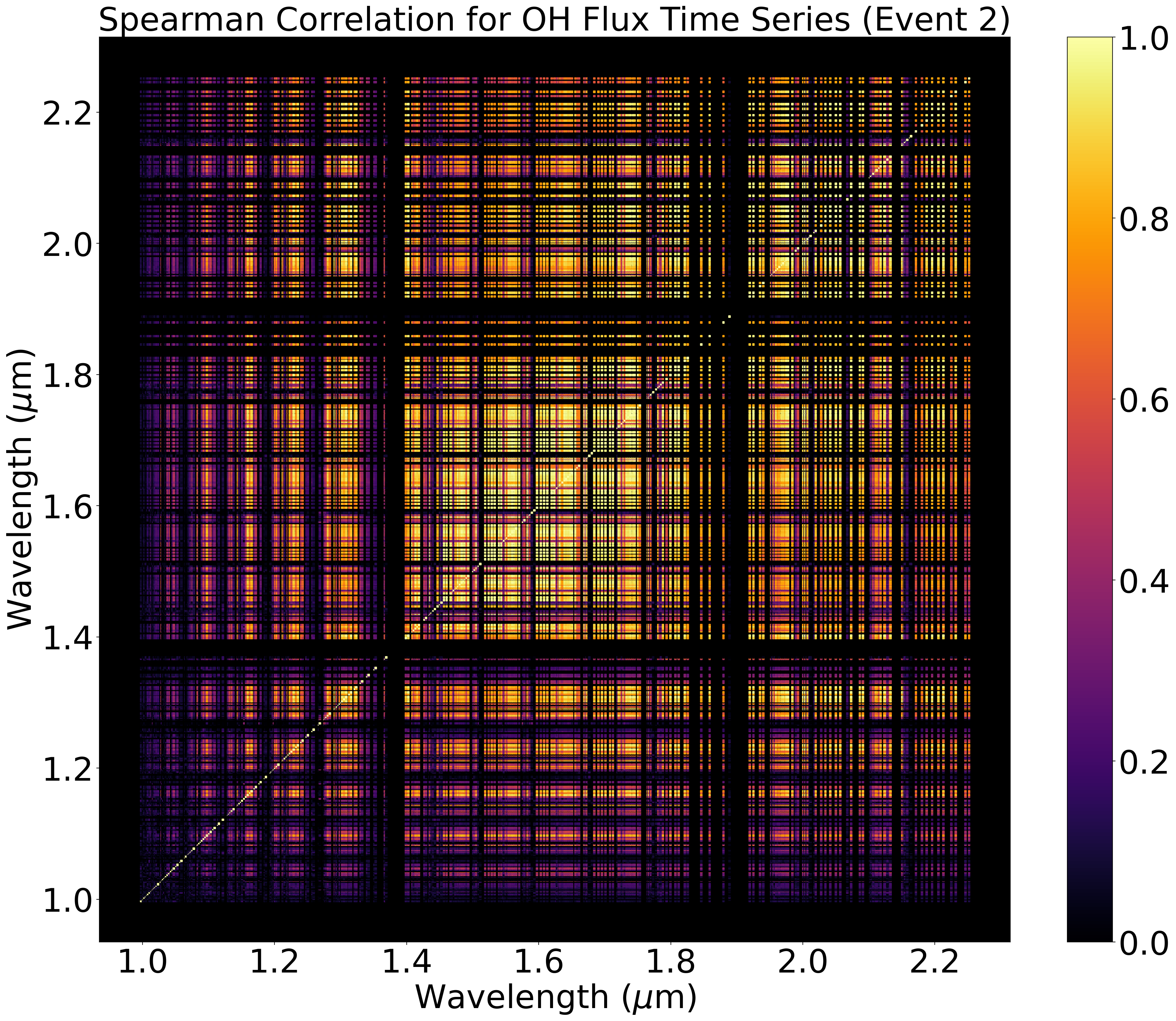}
\caption{Pearson (left) and Spearman (right) correlation heatmaps for the OH flux time series during Event 1 (top) and 2 (bottom). Every data point is the correlation between two different OH flux time series with auto-correlation on the diagonal. The background of the plot is black for easier readability. The low correlations (0.0-0.5) were concentrated in the bluest lines ($1-1.2\mu$m) with the redder lines ($1.8-2\mu$m) also having low correlations in Event 1. However, the latter was the opposite for Event 2, which recorded high correlations (0.5-1), meaning our data was better measured. Overall, a majority of the lines were highly correlated, especially within $1.4-1.8\mu$m.}
\label{fig:correlations}
\end{figure}

Since each OH line goes through the same physical processes, well-measured flux should contain temporal ``families" of OH lines. That is to say, a majority of the OH line flux contributions should be highly correlated. These families are related to similar upper vibrational and rotational transition levels \citep{2023JGRD..12838275N}. We calculated the Pearson and Spearman correlations for each pair of lines (481x481), and detected dozens of highly correlated lines belonging to the same family. From $1-1.3\mu$m (near $J$ band), `dark regions' appearing from poor fitting only reveal several lines belonging to the main family. From $1.4-2.2\mu$m (near $H$ and $K$ bands), dozens of lines belonging to the main family appeared due to excellent fitting. Figure \ref{fig:correlations} describes the result more.

\section{Suggested Machine Learning Approaches\label{suggested_machine_learning}}

Astronomy has made great use of artificial intelligence and machine learning over the past decade because of its accessibility and predictive power. Neural networks, a type of machine learning model that can host millions of parameters, train on an abundance of data to map an input to an output. These models' complexities increase for specific tasks, such as long-short-term-memory (LSTM) neural networks or transformers \citep{2019_lstm,2017_transformer}, which can be used for time series data; these models may be another powerful solution. While a vanilla neural network considers inputs independently, LSTMs and transformers consider inputs sequentially by implementing a feedback mechanic to better learn time-varying features. They have found success in many applications such as speech-to-text, video analysis, and text generation such as GPT-3 \citep{2020_gpt3}. These models are well suited for our problem of predicting OH contributions to the infrared sky. Although a variety of data sets exists, such as from the Network for the Detection of Mesospheric Change (\href{https://ndmc.dlr.de/}{NDMC}), the VLT OH study from \cite{noll23}, or SABER radiometer \citep{1999SPIE.3756..277R}, we prefer building a model for Maunakea and using data sets similar to ours.

Machine learning models can also generate data, such as realistic high-resolution spectra, using variational autoencoders (VAE) or other encoder/decoder machine learning algorithms \citep{2013_vae}. VAEs are the neural network complement to principal component analysis (PCA), which makes it a natural follow-up from PCA since it has been used for sky subtraction. VAE is a dimensionality reduction technique that reduces a large dimensional space into a smaller dimensional latent space by learning an encoder and inverts the smaller dimensional latent space to the original data space using a decoder. Overall, their goal is to reconstruct the original input by compressing the data to a smooth manifold. These models also regularize the latent space to be normally distributed, which makes the latent space smoothly vary. If the VAE is able to sufficiently reconstruct the inputs, then the learned latent space must be an efficient representation of the input data, which could provide a more elegant solution to our problem. Since all the OH lines closely vary to a high degree, performing the same analysis on 481 lines seems redundant. Building a model that captures all that variance and sampling from that latent space vastly decreases that bottleneck with the benefit of capturing the variation along the entire spectrum.

\clearpage
\bibliography{sample631}{}
\bibliographystyle{aasjournal}

\end{document}